\documentclass{emulateapj}
\usepackage{color}
\usepackage{natbib}
\usepackage{amsmath}
\usepackage{epsfig}
\usepackage{times}

\def\be{\begin{eqnarray}}   \def\ee{\end{eqnarray}}
\def\ben{\begin{eqnarray*}} \def\een{\end{eqnarray*}}
\def\sec#1{Section~\ref{sec:#1}}
\def\fig#1{Figure~\ref{fig:#1}}
\def\tab#1{Table~\ref{tab:#1}}
\def\equ#1{Equation~(\ref{equ:#1})}

\newcommand{\Gyr}{\,{\rm Gyr}}
\newcommand{\kpc}{\,{\rm kpc}}

\newcommand{\kms}{\,{\rm km}\,{\rm s}^{-1}}
\newcommand{\kmskpc}{\,{\rm km}\,{\rm s}^{-1}\,{\rm kpc}^{-1}}

\definecolor{grey}{rgb}{0.35,0.35,0.35}
\begin{document}
\submitted{}
\title {Kinematic modelling of the Milky Way using the RAVE and GCS stellar surveys}
\author{S. Sharma   \altaffilmark{1},
J. Bland-Hawthorn \altaffilmark{1},
J. Binney         \altaffilmark{2},
K. C. Freeman     \altaffilmark{3},
M. Steinmetz      \altaffilmark{4},
C. Boeche         \altaffilmark{5},
O. Bienaym{\'e}   \altaffilmark{6},
B. K. Gibson      \altaffilmark{7},
G. F. Gilmore     \altaffilmark{8},
E. K. Grebel       \altaffilmark{5},
A. Helmi          \altaffilmark{9},
G. Kordopatis     \altaffilmark{8},
U. Munari         \altaffilmark{10},
J. F. Navarro     \altaffilmark{11},
Q. A. Parker      \altaffilmark{12},
W. A. Reid        \altaffilmark{12},
G. M. Seabroke    \altaffilmark{13},
A. Siebert        \altaffilmark{6},
F. Watson         \altaffilmark{14},
M. E. K. Williams \altaffilmark{4},
R. F. G. Wyse     \altaffilmark{15},
T. Zwitter        \altaffilmark{16}}
\altaffiltext{1}{Sydney Institute for Astronomy, School of Physics, University of Sydney, NSW 2006, Australia}
\altaffiltext{2}{Rudolf Pierls Center for Theoretical Physics, University of Oxford, 1 Keble Road, Oxford OX1 3NP, UK}
\altaffiltext{3}{RSAA Australian National University, Mount Stromlo Observatory, Cotter Road, Weston Creek, Canberra, ACT 72611, Australia}
\altaffiltext{4}{Leibniz Institut f\" ur Astrophysik Potsdam (AIP), An der Sterwarte 16, D-14482 Potsdam, Germany}
\altaffiltext{5}{Astronomisches Rechen-Institut, Zentrum f\"ur Astronomie der Universit\"at Heidelberg, D-69120 Heidelberg, Germany}
\altaffiltext{6}{Observatoire astronomique de Strasbourg, Universit\'e de Strasbourg, CNRS, UMR 7550, Strasbourg, France}
\altaffiltext{7}{Jeremiah Horrocks Institute for Astrophysics \& Super-computing, University of Central Lancashire, Preston, UK}
\altaffiltext{8}{Institute of Astronomy, University of Cambridge, Madingley Road, Cambridge CB3 0HA, UK}
\altaffiltext{9}{Kapteyn Astronomical Institute, University of Groningen, Postbus 800, 9700 AV Groningen, Netherlands}
\altaffiltext{10}{INAF - Astronomical Observatory of Padova, 36012 Asiago
(VI), Italy}
\altaffiltext{11}{University of Victoria, P.O. Box 3055, Station CSC, Victoria, BC V8W 3P6, Canada}
\altaffiltext{12}{Department of Physics and Astronomy, Macquarie University, Sydney, NSW 2109, Australia}
\altaffiltext{13}{Mullard Space Science Laboratory, University College London, Holmbury St Mary, Dorking, RH5 6NT, UK}
\altaffiltext{14}{Australian Astronomical Observatory, PO Box 296, Epping, NSW 1710, Australia}
\altaffiltext{15}{Johns Hopkins University, 3400 N Charles Street, Baltimore, MD 21218, USA}
\altaffiltext{16}{Faculty of Mathematics and Physics, University of
Ljubljana, Jadranska 19, Ljubljana, Slovenia}
\begin{abstract}
We investigate the kinematic parameters of the Milky Way disc using
the Radial Velocity (RAVE) and Geneva-Copenhagen (GCS) stellar
surveys.  We do this by fitting a kinematic model to the data taking
the selection function of the data into account. For stars in the GCS
we use all phase-space coordinates, but for RAVE stars we use only
$(l,b,v_{\rm los})$. Using Markov Chain Monte Carlo (MCMC) technique,
we investigate the full posterior distributions of the parameters
given the data.  We investigate the `age-velocity dispersion' relation
(AVR) for the three kinematic components ($\sigma_R,\sigma_{\phi},
\sigma_z$), the radial dependence of the velocity dispersions, the
Solar peculiar motion ($U_{\odot},V_{\odot}, W_{\odot} $), the
circular speed $\Theta_0$ at the Sun and the fall of mean
azimuthal motion with height above
the mid-plane.  We confirm that the Besan\c{c}on-style Gaussian model
accurately fits the GCS data, but fails to match the details of the
more spatially extended RAVE survey.  In particular, the Shu
distribution function (DF) handles non-circular orbits more accurately and
provides a better fit to the kinematic data.  The Gaussian
distribution function not only fits the data poorly but systematically
underestimates the fall of velocity dispersion with radius.
The radial scale length of the velocity dispersion profile of the
thick disc was found to be smaller than that of the thin
disc. We find that correlations exist between a number of
parameters, which highlights the importance of doing 
joint fits.  
The large size of the RAVE survey,  
allows us to get precise values for most parameters. 
However, large systematic uncertainties remain,
especially in $V_{\odot}$ and $\Theta_0$. 
We find that, for an
extended sample of stars, $\Theta_0$ is underestimated by as
much as $10\%$ if the vertical
dependence of the mean azimuthal motion is neglected. 
Using a simple model for vertical dependence of kinematics,  
we find that it is possible to match the Sgr A* proper
motion without any need for $V_{\odot}$ being larger than
that estimated locally by surveys like GCS.

\end{abstract}

\section{Introduction}

Understanding the origin and evolution of disc galaxies
is one of the major goals of modern astronomy.
The disc is a prominent feature of late type galaxies like the
Milky Way. As compared to distant galaxies, for which
one can only measure the gross
properties, the Milky Way offers the opportunity to study the
disc in great detail. For the Milky Way, we can determine
6-dimensional phase space information, combined with photometric
and stellar parameters, for a huge sample of stars.
This has led to large observational programs to catalog
the stars in the Milky Way in order to compare them with
theoretical models.

The Milky Way stellar system is broadly composed of four distinct parts
although in reality there is likely to be considerable overlap between
them: the thin disc,
the thick disc, the stellar halo and the bulge. In this
paper, we mainly concentrate on understanding the disc
components which are the dominant stellar populations.

In the Milky Way, the thick disc was originally
identified as the second
exponential required to fit vertical star counts
\citep{1983MNRAS.202.1025G,1993ApJ...409..635R,2008ApJ...673..864J}.
Thick discs are also  ubiquitous features of late
type galaxies \citep{2006AJ....131..226Y}.
But whether the thick disc is a separate component with a distinct
formation mechanism is highly debatable and a difficult
question to answer.

Since the \citet{1983MNRAS.202.1025G} result, various attempts have been made
to characterize the thick disc.  Some studies suggest that thick disc stars
have distinct properties: they are old and metal poor
\citep{2000AJ....119.2843C} and $\alpha$ enhanced
\citep{1998A&A...338..161F,2005A&A...433..185B,2003A&A...410..527B}.
\citet{2008ApJ...673..864J} fit the SDSS star counts using a two-component
model and find that the thick disc has a larger scale-length than the thin
disc.  In contrast, \citet{2012ApJ...753..148B} using a much smaller sample
of SDSS and SEGUE stars
find the opposite when they associate the thick disc with the
$\alpha$-enhanced component.  Finally, the idea of a separate thick disc has
recently been challenged.  \citet{2009MNRAS.399.1145S,2009MNRAS.396..203S}
argued that chemical evolutionary models with radial migration and mixing can
replicate the properties of the thick disc \citep[see also][who explore
radial mixing using N-body simulations]{2011ApJ...737....8L}.
\citet{2008ApJ...684..287I} do not find the expected separation between
metallicity and kinematics for F, G stars in the SDSS survey, and
\citet{2012ApJ...751..131B,2012ApJ...755..115B} argue that the thick disc is
a smooth continuation of the thin disc.

Opinions regarding the formation of a thick disc are equally divided.
Various mechanisms have been proposed: accretion of stars from disrupted
galaxies \citep{2003ApJ...591..499A}, heating of discs by minor mergers
\citep{1993ApJ...403...74Q,2008ApJ...688..254K,2009ApJ...700.1896K,2008MNRAS.391.1806V,2011A&A...525L...3D},
radial migration of stars
\citep{2009MNRAS.399.1145S,2009MNRAS.396..203S,2011ApJ...737....8L}, a
gas-rich merger at high redshift \citep{2004ApJ...612..894B}, and
gravitationally unstable disc evolution \citep{2009ApJ...707L...1B}, inter
alia.  Recently, \citet{2012ApJ...754...48F} have suggested that the thick
disc can form without secular heating, mainly because stars forming at higher
redshift had a higher velocity dispersion.  Another possibility,
proposed by \citet{2010MNRAS.408..783R}, is misaligned angular momentum of
in-falling gas.  How the angular momentum of halo gas becomes misaligned is
described in \citet{2012ApJ...750..107S}.  However,
\citet{2013MNRAS.428.1055A} and \citet{2012MNRAS.423.1544S} suggest that
misaligned gas can destroy the discs.

The obvious way to test the different thick disc theories is to compare the
kinematic and chemical abundance distributions of the thick disc stars with
those of different models.  Since, the thin and thick disc stars strongly
overlap in both space and kinematics, it is difficult to separate them using
just position and velocity.  To really isolate and study the thick disc, one
needs a tag that stays with a star throughout its life. Age is a possible tag
but it is difficult to get reliable age estimates of stars. Chemical
composition is another promising tag that can be used, but this requires high
resolution spectroscopy of a large number of stars. In the near future, surveys
such as GALAH using the HERMES spectrograph \citep{2008ASPC..399..439F} and
the Gaia--ESO survey using the FLAMES spectrograph \citep{2012Msngr.147...25G}
should be able to fill this void.  In our first analysis, we restrict
ourselves to a differential kinematic study of the disc components.  We plan
to treat the more difficult problem of chemo-dynamics in future.

The simplest way to describe the kinematics of the Milky Way stars of the
Solar neighborhood is by assuming Gaussian velocity distributions with some
pre-determined orientation of the principal axes of the velocity ellipsoid.
Then if a single component disc is used, only three components of velocity
dispersion and the mean azimuthal velocity $\overline{v_{\phi}}$ need be
known. If a thick disc is included, one requires five additional parameters, one
of them being the fraction of stars in the thick disc.  If stars are sampled
from an extended volume and not just the Solar neighborhood, then one needs
to specify the radial dependence of the dispersions.

The velocity dispersion of a disc stellar population is known to increase
with age, so one has to adopt an age velocity-dispersion
relation.
Discs heat because a cold, thin disc occupies a very small fraction of phase
space, and fluctuations in the gravitational field cause stars to
diffuse through phase space to  regions of lower phase-space density. The
fluctuations arise from several sources, including giant molecular clouds,
spiral arms, a rotating bar, and halo objects that come close to the disc.
One approach to computing the consequences of these processes is N-body
simulation, but stellar discs are notoriously tricky to simulate accurately,
with the consequence that reliable simulations are computationally costly.
In particular, they are too costly for it to be feasible to find a simulation
that provides a good fit to a significant body of observational data.
Instead we characterize the properties of the Milky Way disc
by fitting a suitable analytical formula. The formula
summarizes large amounts of data but its usefulness
extends beyond this. The formula is traditionally taken to be a power law in age \citep[although
see][]{1993A&A...275..101E,2001ASPC..230...87Q,2007MNRAS.380.1348S}.
The exponents
$\beta_R$, $\beta_{\phi}$ and $\beta_z$ of these power laws
may not be the same for all three components.
The ratio $\sigma_z/\sigma_R$ and the
values of $\beta_R$, $\beta_{\phi}$ and $\beta_z$ are useful for
understanding the physical processes responsible for heating the
disc \citep[e.g.][]{2013seg..book..259B,2013ApJ...769L..24S}.

The first generation of stellar population models characterized the
density distribution of stars using photometric surveys.
\citet{1980ApJ...238L..17B,1980ApJS...44...73B,1984ApJS...55...67B}
assumed an exponential disc with
magnitude-dependent scale heights.
An evolutionary model using
population synthesis techniques was presented by
\citet{1986A&A...157...71R}.
Given a star formation rate (SFR) and an initial
mass function (IMF), one calculates the resulting stellar populations
using theoretical evolutionary tracks. The important step
forward was that the properties of the disc, like scale height,
density laws and
velocity dispersions, were assumed to be a function of age
rather than being color-magnitude dependent terms.
\citet{1987A&A...180...94B} later introduced dynamical self-consistency to
link disc scale and vertical velocity dispersions via the
gravitational potential.
\citet{1997A&A...320..428H,1997A&A...320..440H} further
improved the constraints on SFR and IMF of the disc.
The present state of the art is described in
\citet{2003A&A...409..523R} and is known as the Besan\c{c}on model. Here,
the disc is constructed from a set of isothermal populations that
are assumed to be in equilibrium. Analytic functions for the density
distribution, age/metallicity relation and IMF are provided
for each population. A similar scheme is also used by the
codes TRILEGAL \citet{2005A&A...436..895G} and
{\sl Galaxia} \citep{2011ApJ...730....3S}.

There is a crucial distinction between kinematic and dynamical models.  In a
kinematic model, one specifies the stellar motions independently at each
spatial location, and the gravitational field in which the stars move plays
no role. In a dynamical model, the spatial density distribution of stars and
their kinematics are self-consistently linked by the potential, under the
assumption that the system is in steady state. If one has expressions for
three constants of stellar motion as functions of position and velocity,
dynamical models are readily constructed via Jeans' theorem.
\cite{2012MNRAS.426.1328B} provides an algorithm for evaluating approximate
action integrals, and has used these to fit dynamical models to the GCS data
\citep{2012MNRAS.426.1328B}.  \cite{2014MNRAS.tmp..254B} have confronted the predictions of
the best of these models with RAVE data and shown that the model is remarkably, but not
perfectly, successful.  Our approach is different in two key respects: we fit
kinematic rather than dynamical models, and we avoid adopting distances to, or
using proper motions of, RAVE stars.

Large photometric surveys such as DENIS \citep{1999A&A...349..236E}, 2MASS
\citep{2006AJ....131.1163S} and SDSS \citep{2009ApJS..182..543A} provide the
underpinning for all Galaxy modelling efforts.  The SDSS survey
has been used to provide an empirical model of the Milky Way stars
\citep{2008ApJ...673..864J,2008ApJ...684..287I,2010ApJ...716....1B}.  The
Besan\c con model was fitted to the 2MASS star counts, and its photometric
parameters have been more thoroughly tested than its kinematic parameters
because kinematic data for a large number of stars was not available when the
model was constructed,

The Hipparcos satellite \citep{1997A&A...323L..49P} and the UCAC2 catalog
\citep{2004AJ....127.3043Z} provided proper motions and parallaxes for $\sim
10^5$ stars in the Solar neighborhood. \citet{1998MNRAS.298..387D} used the
Hipparcos data to study stellar kinematics as a function of color. They also
determined the Solar motion with respect to the LSR and the axial ratios of
the velocity ellipsoid. \citet{2000MNRAS.318..658B} also using Hipparcos
stars found the velocity dispersion to vary with function of age as
$\tau^{0.33}$. More recently, \citet{2009MNRAS.397.1286A} using data from a
new reduction of the Hipparcos mission estimated the Solar motion and the AVR
for all three velocity components.  The AVR is assumed to be a power law with
exponents $\beta_R,\beta_{\phi}$ and $\beta_z$ for the three velocity
components in the galactocentric cylindrical coordinate system.  They found
$(\beta_R,\beta_{\phi},\beta_z)=(0.30,0.43,0.44)$. They also investigated the
star formation rate (SFR) and found it to be declining from past to present.
However, a degeneracy exists between the SFR and the slope of the IMF
\citep{1997A&A...320..440H}, and constraining both of them together is
challenging.

The GCS survey \citep{2004A&A...418..989N} combined the Hipparcos and
Tycho-2 \citep{2000A&A...355L..27H} proper motions with radial
velocity measurements and Str\"omgren photometry to create a kinematically unbiased
sample of 16682 F and G stars in the Solar neighborhood. The data
contains full 6D phase space information along with estimates
of ages. The temperature, metallicity and ages were further improved
by \citet{2007A&A...475..519H} and distances and kinematics were
improved by \citet{2009A&A...501..941H} using revised Hipparcos
parallaxes. They investigated the AVR and found $(\beta_R,\beta_{\phi},\beta_z)=(0.39,0.40,0.53)$ which are at odds with
\citet{2009MNRAS.397.1286A}.
\citet{2011A&A...530A.138C} used the infrared flux method to
improve the temperature, metallicity and age estimates for the GCS survey.
The uncertainty in estimated ages is an ongoing concern
for studies that attempt to derive the AVR directly from the GCS data.

With the advent of large spectroscopic surveys like RAVE
\citep{2006AJ....132.1645S} and SDSS/SEGUE \citep{2009AJ....137.4377Y}, we
now have the radial velocity and stellar parameters for a
large number of stars to beyond the Solar neighborhood.
\citet{2012ApJ...755..115B,2012ApJ...751..131B,2012ApJ...753..148B}
used SDSS/SEGUE to fit the spatial distributions of mono-abundance populations
by double exponentials. They showed that
the vertical velocity dispersion declines exponentially with radius
but varies little in $z$. Finally, they argue
that the thick disc is a continuation of the
thin disc rather than a separate entity.

The RAVE survey has also been used to study the stellar
kinematics of the Milky Way disc.
\citet{2012A&A...547A..70P,2012A&A...547A..71P} study the
velocity dispersion and mean motion of the thin and thick disc
stars in the $(R,z)$ plane. They use the technique of singular
value decomposition to compute the moments of the velocity
distribution. Their analysis clearly shows that velocity dispersions
fall as a function of distance from the Galactic Center.
\citet{2013MNRAS.436..101W} explored the kinematics using red clump
stars from RAVE and found complex structures in velocity space.
A detailed comparison with the prediction from the code {\sl Galaxia}
was done, taking
the selection function of RAVE into account. The trend of
dispersions in the $(R,z)$ plane showed a good match with the model.
However, the mean velocities showed significant differences.
\citet{2013A&A...553A..19B} studied the relation between
kinematics and the chemical abundances of stars. By computing
stellar orbits they deduced the maximum vertical distance
$z_{\rm max}$ and eccentricity $e$ of stars. Next they studied
the chemical properties of stars by binning them in the
$(z_{\rm max},e)$ plane. They found that stars with $z_{\rm max}<1\,$kpc
and $0.4<e<0.6$ have two populations with distinct chemical
properties, which hints at radial migration. \citet{2014MNRAS.tmp..254B} used full
six-dimensional information for RAVE stars to fit a
Gaussian model to velocities in the $(v_R,v_z)$ plane. They studied how the
orientation and shape of the velocity ellipsoid varies with location in the
Galaxy, and provided analytic fits to the highly non-Gaussian distributions
of $v_\phi$. They also compared the observed kinematics of stars in different
spatial bins with the predictions of a full dynamical model that had been
fitted to the GCS data.

Stellar kinematics allow us to measure the peculiar motion
($U_{\odot},V_{\odot},W_{\odot}$) of the Sun with respect to the local
standard of rest (LSR), and also the speed of the LSR (in other words, the
circular speed at the location of Sun, $\Theta_0=v_{\rm c}(R_{0})$).
There have been as many determinations of these as there have been new data,
one of the earliest being ($U_{\odot},V_{\odot},W_{\odot})=(9,12,7)$ km
s$^{-1}$ by \citet{1965gast.conf...61D}.  Very precise measurements of these
have been extracted from the Hipparcos proper motions and the Geneva
Copenhagen survey. \citet{1998MNRAS.298..387D} and
\citet{2009MNRAS.397.1286A}, using Hipparcos proper motions, got
($U_{\odot},V_{\odot},W_{\odot})=(9.96\pm0.33,5.25\pm0.54,7.07\pm0.37)$ km
s$^{-1}$.  A revision of $V_{\odot}$ was suggested by
\citet{2010MNRAS.401.2318B} and \citet{2010MNRAS.402..934M}.  Later
\citet{2010MNRAS.403.1829S} explained why the previous estimates, which used
colors as a proxy for age, gave incorrect results. Using a chemo-dynamical model
calibrated on GCS data, they found
($U_{\odot},V_{\odot},W_{\odot})=(11.1\pm0.72,12.24\pm0.47,7.07\pm0.36)$ km
s$^{-1}$.  \citet{2012MNRAS.427..274S} described a model-independent method
and suggests that $U_{\odot}$ could be as high as $14\kms$.  As further
evidence of an unsettled situation, \citet{2012ApJ...759..131B} find from
a sample of 3500 APOGEE stars  $v_{\rm
c}=218\pm 6$, $V_{\odot}=26\pm 3$ and $U_{\odot}=10.5$ km s$^{-1}$ and also
suggest a revision of the LSR reference frame.

\begin{table*}
\caption{\label{tab:geometry} 
Geometry of  stellar components. The formulas used are from
\citet{2003A&A...409..523R}. Note, $(R,\theta,z)$ are the
  coordinates in the galactocentric cylindrical coordinate system and
  $a^2=R^2+\frac{z^2}{\epsilon(\tau)^2}$ (for the
  thin disc). } 
\begin{tabular}{lllll} \hline 
component & age $\tau$ & IMF $\xi(m|\tau)\propto m^{\alpha}$ & density law $\rho(R,z)$ & \\
\hline\hline
Thin disc & $<0.15$ Gyr& $\alpha=1.6$ for $m< 1 M_{\odot}$  & $\propto \exp(-(a/h_{R+})^2)-\exp(-(a/h_{R-})^2)$ & $h_{R+}$ = 5 kpc, $h_{R-}$ = 3 kpc\\
& & $\alpha=3.6$ for $m> 1 M_{\odot}$ & \\
  & 0.15-10 Gyr& &
$\propto  \exp(-(0.5^2+\frac{a^2}{h_{R+}^2})^{0.5})-\exp(-(0.5^2+\frac{a^2}{h_{R-}^2})^{0.5})$ & $h_{R+}$ = 2.53 kpc, $h_{R-}$ = 1.32 kpc\\ 
Thick disc & 11 Gyr & $\alpha=0.5$ & $\propto \exp{(-R/h_{R})} 
  \left(1-\frac{1/h_{z}}{x_l (2.+x_l/h_{z})} z^{2}\right)$ if $|z|\leq x_l$ & $h_{R+}$ = 2.5 kpc, $h_{z}$ = 0.8 kpc\\
 &  &  & $\propto \exp{(-R/h_{R})}\frac{\exp(x_l/h_z)}{1+x_l/2h_z}\exp({-\frac{|z|}{h_{z}}})$ \ \ if $|z| > x_l$  & $x_l=0.4$ kpc
  \\ \hline
\end{tabular}
\end{table*}

In this paper, we refine the
kinematic parameters of the Milky Way, using first a simple
model based on Gaussian velocity distributions, and then
a model based on the Shu distribution function (DF).
We explore the age-velocity dispersion relation, the radial
gradient in dispersions, the Solar motion and the circular
speed. A full exploration of this parameter space using
Markov chain Monte Carlo (MCMC) techniques has not been done before,
even for a sample as small as the GCS.

The RAVE survey contains giants and dwarfs in roughly equal proportions, and
it is hard to determine distances to giants. Moreover, many RAVE stars are
sufficiently distant for the errors in their available, ground-based, proper
motions to give rise to errors in their tangential velocities that far exceed
the small ($\sim1\kms$) errors in their line-of-sight velocities. Hence we
choose not to use either distances or proper motions. Instead we marginalize
over these variables in addition to mass, age, and metallicity.  When the
velocity distribution is Gaussian, the marginalisation over tangential
velocity can be done analytically, in general for other
models, e.g., Shu DF models, the marginalisation has to be
done numerically, and it is computationally expensive.

\citet{2012ApJ...759..131B} recently used a similar procedure to fit models
to 3500 APOGEE stars, but they did not investigate the AVR, and considered
only Gaussian models. In this paper, we fit a kinematic model to 280,000 RAVE
stars taking full account of RAVE's photometric selection function.  To
handle the large data size, we introduce two new MCMC model-fitting
techniques. Our aim is to encapsulate in simple analytical models the main
kinematic properties of the Milky Way disc. Our results should be useful for
making detailed comparison with simulations.

The paper is organized as follows. In \S 2, we introduce the
analytic framework employed for modelling. In \S 3, we describe
the data that we use and its selection functions. In \S 4,
we describe MCMC model-fitting techniques employed here. In \S 5,
we present our results and discuss their implications in \S 6.
Finally, in \S 7 we summarize our findings and look forward to the next
stages of the project.

\section{Analytic framework for modelling the galaxy}\label{sec:framework}
We first describe the analytic framework used to model
the Galaxy \citep{2011ApJ...730....3S}.
The stellar content of the Galaxy is modeled as a set of distinct
components: the thin disc, the thick disc, the stellar halo and the bulge.
The distribution function, i.e.,
the number density of stars
as a function of position (${\bf r}$), velocity (${\bf v}$), age ($\tau$),
metallicity ($Z$), and mass ($m$) for each component, is
assumed to be specified a priori as a function
\be
f_j({\bf r,v},\tau,Z,m)
\label{equ:dist_func1}
\ee
where $j$ $(=1,2,3,4)$ runs over components.
The form of $f_j$ that correctly describes all the properties
of the Galaxy and is self-consistent is still an open question.
However, over the past few decades considerable progress has been made
in identifying a working model dependent on a few simple assumptions
\citep{1986A&A...157...71R,1987A&A...180...94B,1997A&A...320..428H,1997A&A...320..440H,2005A&A...436..895G,2003A&A...409..523R}. Our
analytical framework brings together these models as we describe below.

For a given Galactic component, let the stars form at a rate
$\Psi(\tau)$ with a mass distribution
$\xi(m|\tau)$ (IMF) that is a parameterized
function of age $\tau$. Let the present day
spatial distribution of stars $p({\bf r}|\tau)$ be
conditional on age only. Finally, assuming that the
velocity distribution to be $p({\bf v|r},\tau)$ and the
metallicity distribution to be $p(Z|\tau)$, we have
\be
f({\bf r,v},\tau,m,Z) = \frac{\Psi(\tau)}{\langle m\rangle}\xi(m|\tau)p({\bf r}|\tau)p({\bf v|r},\tau)p(Z|\tau) .
\label{equ:dist_func2}
\ee
The functions conditional on age can take
different forms for different Galactic components.
The IMF here is normalized such
that $\int_{m_{\rm min}}^{m_{\rm max}}\xi(m|\tau)dm =1$ and
$\langle m \rangle=\int_{m_{\rm min}}^{m_{\rm max}}m\xi(m|\tau)dm$ is
the mean stellar mass.
The  metallicity distribution  is modeled as a log-normal distribution,
\be
p(Z,|\tau)=\frac{1}{\sigma_{\log Z}(\tau)\sqrt{2\pi}}{\rm exp}\left[-\frac{(\log Z-\log\bar{Z}(\tau))}{(2\sigma_{\log
Z}^2(\tau))}\right],\label{equ:amr}
\ee
the mean and dispersion of which are given by  age-dependent functions
$\bar{Z}(\tau)$ and $\sigma_{\log Z}(\tau)$.
The $\bar{Z}(\tau)$ is widely referred to as the age-metallicity
relation (AMR).
Functional forms for each of the expressions
in \equ{dist_func2} are given in \citet{2011ApJ...730....3S}
\citep[see also][]{2003A&A...409..523R}. 
For convenience 
we reproduce in \tab{geometry} a short description  of the thin 
and thick disc components.
The axis ratio $\epsilon$ of the thin disc is given by
\be
\epsilon(\tau)={\rm Min}\left(0.0791,0.104\left(\frac{\tau/\text{Gyr}+0.1}{10.1}\right)^{0.5}
  \right),
\ee
and this represents the age scale height relation.

\subsection{Kinematic modelling}
Having described the general framework for analytical modelling, 
we now discuss our strategy for the kinematic modelling of the 
Milky Way. Simply put, we want to 
constrain the velocity distribution $p({\bf v|r},\tau)$.
In what follows, we assume that everything 
except for $p({\bf v|r},\tau)$ on the 
right hand side of \equ{dist_func2} is known. 
In the next two subsections we discuss the 
functional forms of the adopted $p({\bf v|r},\tau)$ 
and describe ways to parameterize them. 
Technical details related to fitting such a model 
to observational data are discussed in 
\sec{modtech}.

Although we can supply any functional form for  
$p({\bf v|r},\tau)$ and fit them to data, in 
reality there is much less freedom. 
The spatial density distribution and 
the kinematics are linked to each other 
via the potential. 
Hence, specifying $p({\bf v|r},\tau)$ independently 
lacks self consistency. 
In such a scenario, the accuracy of a pure 
kinematic model depends upon our ability to 
supply functional forms of $p({\bf v|r},\tau)$ 
that are a good approximation to the actual 
velocity distribution of the system.
A proper way to handle 
this problem would be to use  
dynamically self consistent models, but such models 
are still under development and we hope 
to explore them in future. In the meantime, we explore 
kinematic models that provide a reasonable 
approximation to the actual velocity distribution 
and hope to learn from them.

\subsection{Gaussian velocity ellipsoid model}
In this model, the velocity distribution is assumed to be a triaxial
Gaussian,
\begin{eqnarray}
p({\bf v|r},\tau) & = &
\frac{1}{\sigma_{R}\sigma_{\phi}\sigma_{z}(2\pi)^{3/2}}
{\rm  exp}\left[-\frac{v_{R}^2}{2\sigma_{R}^2}\right]
{\rm  exp}\left[-\frac{v_z^2}{2\sigma_z^2}\right] \nonumber \\
& &  \times{\rm exp}\left[-\frac{(v_{\phi}-\overline{v_{\rm \phi}})^2}{2\sigma^2_{\phi}} \right],
\label{equ:veldist}
\end{eqnarray}
where $R,\phi,z$ are cylindrical coordinates.
The $\overline{v_{\phi}}$ is the asymmetric drift and is
given by
{\small
\be
\overline{v_{\phi}}^2(\tau,R)  & = &  v_{\rm c}^2(R)+ \sigma_{R}^2 \nonumber \\
& &   \times\left( \frac{d \ln \rho}{d \ln
R}+ \frac{d \ln \sigma_{R}^2}{d \ln R}+1-\frac{\sigma_{\phi}^2}{\sigma_R^2}+1-\frac{\sigma_z^2}{\sigma_R^2}\right)
\label{equ:stromberg}
\ee
}
This follows from Equation 4.227 in \citet{2008gady.book.....B}
assuming $\overline{v_R\:v_z}=(v_R^2-v_z^2)(z/R)$. This is valid for
the case where the principal axes of velocity ellipsoid are aligned
with the $(r,\theta,\phi)$ spherical coordinate system. If the
velocity ellipsoid is aligned with the cylindrical $(R,\phi,z)$
coordinate system, then $\overline{v_R\:v_z}=0$.
Recent results using the RAVE data suggest that the velocity ellipsoid is aligned
with the spherical coordinates \citep{2008MNRAS.391..793S,2014MNRAS.tmp..254B}.
One can parameterize our ignorance by writing the asymmetric drift as follows:
{\small
\be
\overline{v_{\phi}}^2(\tau,R) & = & v_{\rm c}^2(R)+ \sigma_{R}^2 \left( \frac{d \ln \rho}{d \ln
R}+ \frac{d \ln \sigma_{R}^2}{d \ln
R}+1-k_{\rm ad}^2\right)
\label{equ:stromberg_bovy}
\ee
}
This is the form that is used by \citet{2012ApJ...759..131B}.

The dispersions of the $R,\phi$ and $z$ components of velocity
increase as a function age due to secular heating in the disc,
and there is a radial dependence such that the dispersion
increases towards the Galactic Center. We model these effects after
\citet{2009MNRAS.397.1286A} and \citet{2010MNRAS.401.2318B}
using the functional form
\begin{eqnarray}
\sigma^{\rm thin}_{R,\phi,z}(R,\tau) & = &\sigma_{R,\phi,z,\odot}^{\rm
thin}\exp\left[-\frac{R-R_{0}}{R_\sigma^{\rm thin}}\right]
\nonumber \\ & &
\times\left(\frac{\tau+\tau_{\rm
min}}{\tau_{\rm max}+\tau_{\rm
min}}\right)^{\beta_{R,\phi,z}}
\\
\label{equ:veldisp1}
\sigma^{\rm thick}_{R,\phi,z}(R) & = &\sigma_{R,\phi,z,\odot}^{\rm thick}
{\rm exp}\left[-\frac{R-R_{0}}{R_\sigma^{\rm thick}}\right].
\label{equ:veldisp2}
\end{eqnarray}
The choice of the radial dependence is motivated by the desire to
produce discs in which
the scale height is independent of radius.
For example, under the epicyclic approximation, if $\sigma_z/\sigma_R$
is assumed to be constant, then
the scale height is independent of radius for $R_\sigma=2R_d$
\citep{1982A&A...110...61V,1988A&A...192..117V,2011ARA&A..49..301V}.
In reality there is also a $z$ dependence of velocity dispersions
which we have chosen to ignore in our present analysis.
This means that for
a given mono age population the asymmetric drift is independent
of $z$. However, the velocity dispersion and asymmetric drift
of the combined population of stars are functions of $z$. This
is because the scale height of stars for a given isothermal
population is an increasing function of
its vertical velocity dispersion.

For our kinematic analysis we assume $d\ln \rho/d R=-1/R_d $ 
with $R_d=2.5 \kpc$. 
While this is true for the thick disc adopted by us, for the thin 
disc this is only approximately true (see \tab{geometry}). 
The thin disc with age 
between 0.15 and 10 Gyr is exponential at large $R$ with  
a scale length of 2.53 kpc.

\subsection{Shu distribution function model}

The Gaussian velocity ellipsoid model has its limitations. In particular,
the distribution of $v_\phi$ is strongly non-Gaussian, being
highly skew to low $v_\phi$.

For a two-dimensional disc, a much better approximation to the velocity
distribution is provided by the \citet{1969ApJ...158..505S} distribution
function.  Moreover, the Shu DF, being dynamical in nature, connects the
radial and azimuthal components of velocity dispersion to each other and to
the mean-streaming velocity, thus lowering the number of free parameters in
the model.

Assuming the potential is separable as
$\Phi(R,z)=\Phi_R(R)+\Phi_z(z)$ we can write the
distribution function as
\be
f(E_{R},L_z,E_z)=\frac{F(L)}{\sigma_{R}^2(L_z)}{\rm
exp}\left[-\frac{E_{R}}{\sigma_{R}^2(L_z)}\right] \nonumber \\
\times \frac{{\rm
exp}\left[-(E_z)/(\sigma^2_z(L_z))\right]}{\sigma_z(L_z)\sqrt{2\pi}},
\ee
where $L=Rv_\phi$ is the angular momentum,
\be
E_z=\frac{v_z^2}{2}+\Phi_z(z)
\ee

\be
E_R &= & \frac{1}{2}v_R^2+\Phi_{\rm eff}(R,L_z)-\Phi_{\rm eff}(R_g,L_z)
\nonumber \\
& = & \frac{1}{2}v_R^2+\Delta \Phi_{\rm eff}(R,L_z)
\ee
with
\be
\Phi_{\rm eff}(R,L_z) &= & \frac{L_z^2}{2R^2}+\Phi(R)\simeq
\frac{L_z^2}{2R^2}+v^2_c\ln R \label{equ:effpot}
\ee
being the effective potential.
Let $R_g(L_z)=L_z/v_{\rm c}$ be the radius of a circular orbit with
specific angular momentum $L_z$. In \citet{2012MNRAS.419.1546S} \citep[see also][]{2013ApJ...773..183S} it was shown
that joint distribution of $R$ and $R_g$ can be written as
{\small
\begin{eqnarray}
P(R,R_g)   & = & \frac{(2\pi)^2\Sigma(R_g)}{g(\frac{1}{2a^2})}{\rm
exp}\left[\frac{2 \ln(R_g/R)+1-R_g^2/R^2}{2a^2}\right],
\label{equ:prrl_shu}
\end{eqnarray}
}
where $\Sigma(R)$ is a function that controls the disc's surface density and
\begin{eqnarray}
a & = & \sigma_{R}(R_g)/v_{\rm c} \label{equ:a_shu} \\
g(c) & = & \frac{e^c\Gamma(c-1/2)}{2c^{c-1/2}}.
\end{eqnarray}

We assume $a$ to be  specified as
\begin{eqnarray}
a & = &a_0(\tau) {\rm exp}\left[-\frac{R_g}{R_\sigma}\right]\nonumber \\
& = &\frac{\sigma_{R,\odot}}{v_{\rm c}}\left(\frac{\tau+\tau_{\rm
min}}{\tau_{\rm max}+\tau_{\rm min}}\right)^{\beta_R}
{\rm exp}\left[-\frac{R_g-R_{0}}{R_\sigma}\right]
\end{eqnarray}
and $\sigma_z$ to be  specified as
{\small
\be
\sigma_{z0}(R_g,\tau) & = &\sigma_{z,\odot}\left(\frac{\tau+\tau_{\rm
min}}{\tau_{\rm max}+\tau_{\rm min}}\right)^{\beta_z}
{\rm exp}\left[-\frac{R_g-R_{0}}{R_\sigma}\right].
\ee
}
Now this leaves us to choose $\Sigma(R_g)$. This should be
done so as to produce discs that satisfy the observational
constraint given by $\Sigma(R)$, i.e., an exponential disc (or discs)
with scale length $R_d$. A simple way to do this is
to let
\be
\Sigma(R_g) & = &  \frac{e^{-R_g/R_d}}{2 \pi R_d^2} .
\ee
However, this matches the target surface density only approximately.
A better way to do this is to
use the empirical formula proposed in \citet{2013ApJ...773..183S}
such that
{\footnotesize
\be
\Sigma(R_g) & = & \frac{e^{-R_g/R_d}}{2\pi R_d^2} - \frac{0.00976
a^{2.29}_{0}}{R_d^2}  s\left[\frac{R_g}{(3.74R_d(1+q/0.523)}\right]
\ee
}
where $q=R_d/R_\sigma$ and $s$ is a function of the following form
\be
s(x) & = &k e^{-x/b}((x/a)^2-1),
\ee
with $(k,a,b)=(31.53,0.6719,0.2743)$.
This is the scheme that we employ in this paper.
As in the previous section, we adopt $R_d=2.5$ kpc.

\begin{table}
\caption{\label{tab:tbparam}
Description of model parameters}
\centering
\begin{tabular}{|l|l|} \hline
Model Parameter           & Description \\ \hline
$U_{\odot}$                & Solar motion with respect to LSR \\ \hline
$V_{\odot}$                & Solar motion with respect to LSR \\ \hline
$W_{\odot}$                & Solar motion with respect to LSR \\ \hline
$\sigma_R^{\rm thin}$      & The velocity dispersion at 10 Gyr \\
& Normalization of thin disc AVR (Eq \ref{equ:veldisp1}) \\ \hline
$\sigma_{\phi}^{\rm thin}$  & The velocity dispersion at 10 Gyr \\
& Normalization of thin disc AVR (Eq \ref{equ:veldisp1}) \\ \hline
$\sigma_z^{\rm thin}$      & The velocity dispersion at 10 Gyr \\
& Normalization of thin disc AVR (Eq \ref{equ:veldisp1}) \\ \hline
$\sigma_R^{\rm thick}$     & The velocity dispersion of thick disc (Eq \ref{equ:veldisp2})\\ \hline
$\sigma_{\phi}^{\rm thick}$ & The velocity dispersion of thick disc (Eq \ref{equ:veldisp2})\\ \hline
$\sigma_z^{\rm thick}$     & The velocity dispersion of thick disc (Eq \ref{equ:veldisp2})\\ \hline
$\beta_R$                & The exponent of thin disc AVR (Eq \ref{equ:veldisp1})\\ \hline
$\beta_{\phi}$            & The exponent of thin disc AVR (Eq \ref{equ:veldisp1})\\ \hline
$\beta_z$                & The exponent of thin disc AVR (Eq \ref{equ:veldisp1})\\ \hline
$R_{\sigma}^{\rm thin}$             & The scale length  of the\\
& velocity dispersion profile for thin disc (Eq \ref{equ:veldisp1})\\ \hline
$R_{\sigma}^{\rm thick}$            & The scale length of the \\
& velocity dispersion profile for thick disc (Eq \ref{equ:veldisp2})\\ \hline
$R_{0}$                  & Distance of Sun from the Galactic Center\\ \hline
$\Theta_0$          & The circular speed at Sun \\ \hline
$\alpha_{z}$          & Vertical fall of circular
velocity (Eq \ref{equ:defsvc}) \\ \hline
$\alpha_{R}$          & Radial gradient of circular speed (Eq
\ref{equ:defsvc})\\ \hline
\end{tabular}
\end{table}

\subsection{Model for the potential} \label{sec:potmodel}
\begin{figure}
\centering \includegraphics[width=0.50\textwidth]{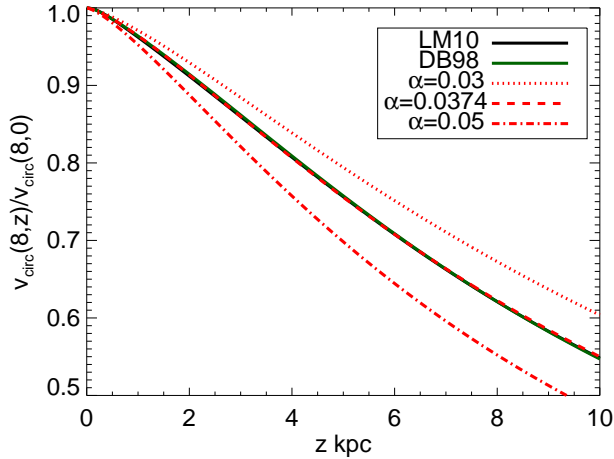}\caption{Circular
  speed as a function of height $z$ above the mid plane for models of the Milky Way consisting of bulge, halo and
disc. The non solid red lines are for the fitting formula with
different values of $\alpha_z$. The larger the $\alpha_z$ the
steeper is the fall of circular speed.
\label{fig:vcirc_z}}
\end{figure}

So far we have described kinematic models in which the
potential is separable in $R$ and $z$. In such cases,
the energy associated with the vertical motion
$E_z$ can be assumed to be the third integral of motion.
In reality, the potential generated by a double exponential
disc is not separable in $R$ and $z$.
For example, the hypothetical circular speed
defined as $\sqrt{R\partial{\Phi(R,z)}/\partial{R}}$ can have
both a radial and a vertical dependence.
We model it as 
\begin{eqnarray}\label{equ:defsvc}
v_{\rm c}(R,z) & = & \sqrt{R \frac{\partial \Phi}{\partial R}}  \nonumber \\
& = & (\Theta_0+\alpha_R (R-R_{\odot}))\frac{1}{1+\alpha_z |z/\kpc|^{1.34}}.
\end{eqnarray}
The parameters $\alpha_R$ and $\alpha_z$
control the radial and vertical dependencies, respectively. The motivation
for the vertical term comes from the fact that the above formula
with $\alpha_z=0.0374$ provides a good fit to the $v_c(R_0,z)$
profile of Milky Way potential by \citet{1998MNRAS.294..429D} as
well as that of \citet{2010ApJ...714..229L} (see \fig{vcirc_z}). Both of them
have bulge, halo and disc components. The former has two double
exponential discs while the later has a Miyamoto-Nagai disc.

To accurately model a system, in which the potential is not separable in
$R$ and $z$, requires a distribution function that
incorporates the third integral of motion in addition to
energy $E$ and angular momentum $L_z$, e.g., distribution
functions based on action integrals $J_r,J_z$ and $L_z$ \citep{2012MNRAS.426.1328B,2010MNRAS.401.2318B}. Converting phase
space coordinates $(x,v)$ to actions integrals is not easy and
techniques to make this possible are under development.
One way to compute the actions is by using the adiabatic
approximation, i.e., conservation
of vertical action  \citep{2011MNRAS.413.1889B,2012MNRAS.419.1546S}.
Using an adiabatic approximation, \citet{2012MNRAS.419.1546S}
extend the Shu DF to three dimensions and model the
kinematics as a function of distance from the plane.
Recently, it has been shown by \citet{2012MNRAS.426.1324B}
that the adiabatic approximation is accurate only close to
the midplane and that much better results are obtained
by assuming the potential to be similar to a Stackel potential.

In this paper, to model systems where the potential is not
separable in $R$ and $z$, we follow a much simpler
approach. The approach is motivated by the fact that,
for realistic galactic potentials,  we expect
the $\overline{v_{\phi}}$ of a single age population to fall
with $z$.
It has been shown by both \citet{2011MNRAS.413.1889B} and
\citet{2012MNRAS.419.1546S} that when
vertical motion is present, in a Milky Way type potential,
the effective potential for
radial motion (see Equation \ref{equ:effpot}) needs to be
modified as the vertical motion
also contributes to the centrifugal potential.
Neglecting this
effect leads to an overestimation of $\overline{v_{\phi}}$.
As one moves away from the plane this effect is expected to
become  more and more important. Secondly,
as shown by \citet{2012MNRAS.419.1546S}, in a given solar
annulus, stars with smaller $R_g$ will have larger vertical
energy and hence larger scale height. This implies that stars with
smaller $R_g$ are more likely to be found at higher $z$,
consequently the $\overline{v_{\phi}}$ should also decrease
with height.

The fall of $\overline{v_{\phi}}$ with height is also
predicted by the Jeans equation for an axisymmetric system
\be
\overline{v_{\phi}}^2(R,z) & = & \left[R\frac{\partial \Phi}{\partial R}\right] +\sigma_R^2
\left[1-\frac{\sigma_{\phi}^2}{\sigma_R^2}  
 +\frac{\partial{\ln(\rho\sigma_R^2)}}{\partial{\ln
R}}\right] \nonumber \\
& & + R\left[ \frac{\partial \overline{v_R
v_z}}{\partial z} +\overline{v_R
v_z}\frac{\partial \ln \rho}{\partial z}\right]. \label{equ:jeanaxis}
\ee
The $\overline{v_{\phi}}$ at high
$z$ will be lower both because $R\partial\Phi /\partial R$
is lower and because the term in the third square bracket
decreases with $z$, e.g., assuming $\overline{v_R
v_z}=(\sigma_R^2-\sigma_z^2)z/R$.

For the Gaussian model we simulate the
overall reduction of  $\overline{v_{\phi}}$ with $z$
by introducing a parameterized form for $v_c(R,z)$ as given
by \equ{defsvc} in \equ{stromberg}. Given this prescription
we expect  $\alpha_z>0.03744$, so as to account for effects other
than that involving the first term in \equ{jeanaxis}.
In reality, the
velocity dispersion tensor ${\bf \sigma}^2$ will have a much
more complicated dependence on $R$ and $z$ than what
we have assumed, e.g.,  we assume
that $\sigma_{R,\phi,z}$ only has an $R$ dependence which
is given by an exponential form.

For the Shu model we replace $v_c$ in \equ{a_shu} by
the form in \equ{defsvc}. The idea again is to model
the fall of  $\overline{v_{\phi}}$ with $z$.
However, the prescription breaks the dynamical
self-consistency of the model and turns it into a
fitting formula. In reality, the $\overline{v_{\phi}}$ may
not exactly follow the functional form for the vertical
dependence predicted by our model, but is better than
completely neglecting it.

\subsection{Models and parameters explored}
We now give a description of the parameters and models that
we explore. We investigate up to 18 parameters
(see \tab{tbparam} for a summary). These are
the Solar motion $(U_{\odot},V_{\odot},W_{\odot})$, the logarithmic
slopes of age-dispersion relations $(\beta_R,\beta_{\phi},\beta_z)$,
the scale lengths of radial dependence of velocity dispersions
$(R_\sigma^{\rm  thin},R_\sigma^{\rm thick})$, the velocity dispersions at $R=R_{0}$
of the thin disc
$(\sigma_{\phi}^{\rm thin},\sigma_z^{\rm  thin},\sigma_R^{\rm thin})$
and of the thick disc
$(\sigma_{\phi}^{\rm  thick},\sigma_z^{\rm  thick},\sigma_R^{\rm
thick})$; for simplicity the subscript $\odot$ is dropped here.
The Gaussian models are
denoted by GAU whereas models based on the Shu DF
are denoted by SHU.
For models based on the Shu DF, the azimuthal motion
is coupled to the radial motion, hence $\beta_{\phi}$,
$\sigma_{\phi}^{\rm thin}$ and $\sigma_{\phi}^{\rm thick}$ are not
required.
When $\Theta_0$ is fixed, we assume its
value to be $226.84\kms$.
In some cases, we also keep the parameters
$\beta_z$ and $R_\sigma^{\rm thin}$ fixed. While reporting the results we
highlight the fixed parameters using the magenta color.

In our analysis the distance of the Sun from the galactic center,
$R_0$, is assumed to be 8.0 kpc.  To gauge the sensitivity of our
results to $R_0$, we also provide results for cases with $R_0=7.5$
and 8.5 kpc.
The true value of $R_0$ is still debatable
ranging from 6.5 to 9 kpc.
Recent results from studies of orbit of stars
near the Galactic Center give $R_0=8.33\pm0.35$
\citep{2009ApJ...692.1075G}.
The classically accepted value
of $8 \pm 0.5$ kpc is a weighted average given in a review by
\citet{1993ARA&A..31..345R}.
The main reason we keep $R_0$ fixed is as follows.
Given that we do not make use of explicit distances, proper motions
or external constraints like the proper motion of Sgr A*, it is
clear we will not be able to constrain
$R_0$ well, specially if $\Theta_0$ is free.
For example \citet{2010MNRAS.402..934M} using parallax, proper motion
and line of sight velocity of masers in high
star forming regions, show that constraining both $\Theta_0$ and $R_{0}$
independently is difficult.

\section{Observational data and selection functions}
In this paper we analyze data from two surveys, the Radial
Velocity Experiment, RAVE
\citep{2006AJ....132.1645S,2008AJ....136..421Z,2011AJ....141..187S,2013AJ....146..134K}
and the Geneva Copenhagen Survey, GCS
\citep{2004A&A...418..989N,2009A&A...501..941H}.
For fitting theoretical models to data from stellar surveys,
it is important to take into
account the selection biases that were introduced when observing the
stars. This is especially important for spectroscopic surveys which
observe only a subset of the all possible stars
defined within a color-magnitude range. So we also analyze the
selection function for the RAVE and GCS surveys.

\subsection{RAVE survey}
The RAVE survey collected spectra of 482430 stars between April 2004 and
December 2012 and stellar parameters, radial velocity, abundance and
distances have been determined for 425561 stars.  In this paper we used the
internal release of RAVE from May 2012, which consisted of 458412
observations.  The final explored sample after applying various selection
criteria consists of 280128 unique stars.  These data are available in the
DR4 public release \citep{2013AJ....146..134K}, where an extended discussion
of the sample is also presented.

For RAVE we only make use of the $\ell,b$ and $v_{\rm los}$ of stars.
The $I_{\rm DENIS}$ and 2MASS $J-K_s$ colors are used for
marginalization over age, metallicity and mass of stars
taking into account the photometric selection function of RAVE.
We do not use proper motions, or stellar parameters which could in principle
provide tighter constraints, but then one has to worry
about the systematics introduced by their use.
For example,
in a recent kinematic analysis of RAVE stars,
\citet{2013MNRAS.436..101W} found
systematic differences between different
proper motion catalogs like PPMXL \citep{2008A&A...488..401R},
SPM4 \citep{2011AJ....142...15G}
and UCAC3 \citep{2010AJ....139.2184Z}.
As for stellar parameters, although they are reliable,
no pipeline can claim to be free of unknown systematics
specially when working with low signal to noise data.
Hence, as a first step it is  instructive to
work with data that are least ambiguous and then in the next
step check the results by adding more information.
As we will show later, for the types of model that we consider,
even using only $\ell,b$ and $v_{\rm los}$ can
provide  good constraints on the model parameters.

We now discuss the selection function of RAVE.
The RAVE survey was designed to be a magnitude-limited survey in
the $I$ band. This means that theoretically it has one of the simplest
selection functions, but, in practice, for a multitude of reasons,
some biases were introduced. First, the DENIS and
2MASS surveys were not fully available when the survey started. Hence,
the first input catalog (IC1) had stars
from Tycho and SuperCOSMOS. For Tycho stars, $I$ magnitudes
were estimated from $V_{T}$ and
$B_{T}$ magnitudes. On the other hand, the SuperCOSMOS stars had
$I$ magnitudes but an offset was
later detected with respect to $I_{\rm DENIS}$. Later, as DENIS and
2MASS became available, the second input catalog IC2 was created.
With the availability of DENIS, it became possible to have a direct
$I$ mag measurement which was free from offsets like those observed in
SuperCOSMOS.
But DENIS itself had its own problems -- saturation
at the bright end, duplicate entries, missing stripes in the sky, inter alia.
To solve the problem of duplicate entries, the DENIS catalog
was cross-matched with 2MASS to within a tolerance of
1$^{\prime\prime}$. This helped
clean up the color-color diagram of $(I_{\rm DENIS}-K_{\rm 2MASS})$ vs
$(J_{\rm 2MASS}-K_{\rm 2MASS})$ in particular \citep{2007_Seabroke_Thesis}.

Given this history, the question arises how can we compute the selection
function. Since accurate $I$ mag photometry is not available for stars that
are only in IC1, the first cut we make is to select stars from IC2 only.
Then we removed the duplicates-- among multiple observations one of them was
selected randomly.  To weed out stars with large errors in radial velocity,
we made some additional cuts:
\ben
{\rm Signal\ to \ Noise\  STN} > 20 \\
{\rm Tonry-Davis \ Correlation \ Coefficient} > 5.
\een

For brighter magnitudes, $I_{\rm DENIS}<10$, $I_{\rm DENIS}$
suffers from saturation.
One could either get rid of these stars to be more
accurate or ignore the saturation.
In the present analysis we ignore the saturation.
Note, the observed stars in the input catalog are not necessarily
randomly sampled from the IC2. Stars were divided into four bins
in $I_{\rm mag}$ and stars in each bin were randomly selected to observe
at a given time. However it seems later on this division was
not strictly maintained (probably due to the
observation of calibration stars and some extra stars going to
brighter magnitudes).
This means the selection function has to
be computed as a function of $I_{\rm DENIS}$ in much finer bins.
Assuming the DENIS I magnitudes are correct, and the cross-matching is correct,
the only thing that needs to be taken into account is the
angular completeness of the DENIS survey (missing stripes).
To this end, we grid the observed
and IC2 stars in $(\ell,b,I_{\rm DENIS})$ space and compute a probability map.
To grid the angular co-ordinates we use the HEALPIX
pixelization scheme \citep{2005ApJ...622..759G}.
The resolution of HEALPIX is specified by
the number $n_{\rm side}$ and the total number of pixels is given
by $12n_{\rm side}^2$. For our purpose, we use $n_{\rm side}=16$ which
gives a pixel size of 13.42 ${\rm deg}^2$, which is smaller than the
RAVE field of view of 28.3 ${\rm deg}^2$. For magnitudes,
we use a bin size of 0.1 mag, which again is much smaller than
the magnitude range included in each observation. Given the fine
resolution of the probability map, the angular and magnitude
dependent selection biases are adequately handled.
Note, in the range $(225^\circ<\ell<315^\circ)\ \&\ (5^\circ<|b|<25^\circ)$,
a color selection of $(J-K_s)>0.5$ was used to selectively target
giants, and we take this into account in our analysis.

\begin{figure}
\centering \includegraphics[width=0.50\textwidth]{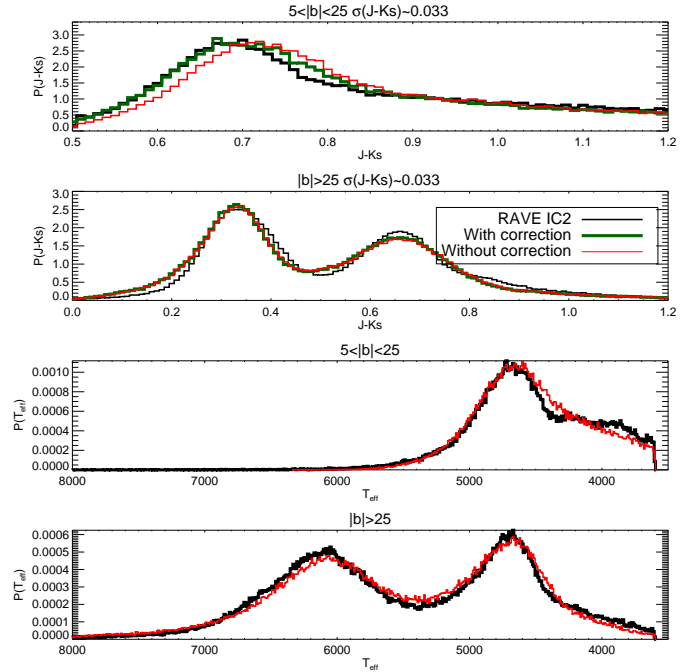}\caption{The
  color and temperature distribution (from DR3 pipeline) of RAVE stars compared with {\sl Galaxia} simulations with properly
matched selection
and statistical sampling. The effect of our new correction formula for the
Schlegel extinction map is also shown.
The results for $|b|<25^\circ$ and $|b|>25^\circ$ are shown
separately. Note, {\sl Galaxia} makes use of Padova isochrones.
\label{fig:extcorr}}
\end{figure}

\citet{1999ApJ...512L.135A} suggest that the \citet{1998ApJ...500..525S} maps
overestimate reddening by a factor of 1.3-1.5 in regions with smooth
extinction $A_V>0.5$, i.e., $E_{B-V} > 0.15$ \citep[see
also][]{2005A&A...435..131C}.  In \fig{extcorr} the color and temperature
distributions of our RAVE stars (black lines) are compared with predictions
from {\sl Galaxia} given the selection above.  At high latitudes (second and
fourth panels) the red model curves agree reasonably well with the black data
curves, but in the top panel ($5^\circ<|b|<25^\circ$) the red model
distribution of $J-K$ colors is clearly displaced to red colours relative to
the data.  The low-latitude temperature distributions shown in the third
panel show no analogous shift of the model curve to lower temperatures, so we have a clear
indication that the model colours have been made too red by excessive
extinction. To correct this problem, we modify the Schlegel $E_{B-V}$
as follows
\be f_{\rm
corr}& = & 0.6+0.2\left(1-{\rm
tanh}\left[\frac{E_{B-V}-0.15}{0.1}\right]\right)
\ee
The formula above reduces extinction by 40\% for high extinction regions;
the transition occurs around $E_{B-V} \sim 0.15$ and is smoothly controlled
by the ${\rm tanh}$ function.  The green curves in the top two panels show
that the proposed correction to Schlegel maps. Although not perfect, the
correction reduces the discrepancy between the model and data for low
latitude stars (top panel) whilst having negligible impact on high-latitude
stars.

The fact that the temperature and color distributions in \fig{extcorr} match
up so well is encouraging, given that we selected on $I_{\rm DENIS}$
magnitude alone. This implies that the spatial distribution of stars
specified by {\sl Galaxia} satisfies one of the necessary observational
constraints.

\begin{figure}
\centering \includegraphics[width=0.48\textwidth]{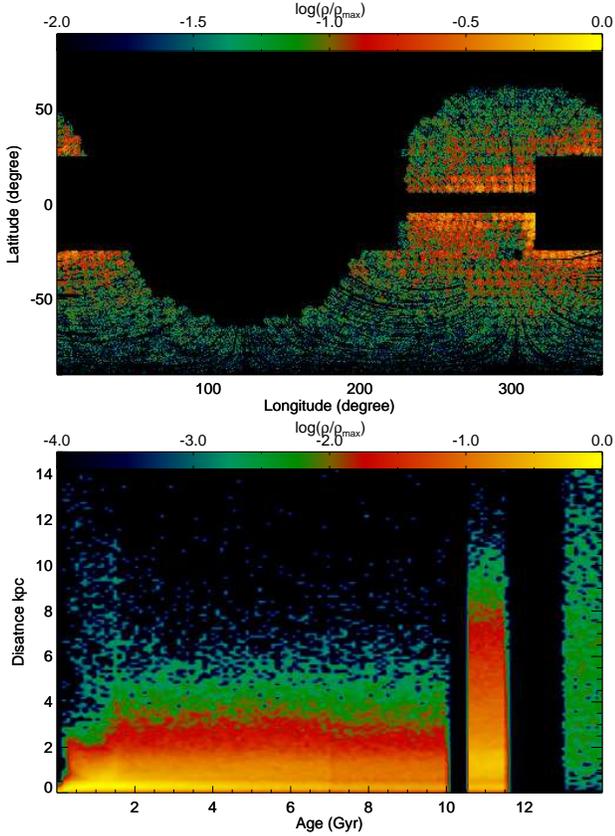}\caption{Probability
  distribution of RAVE stars analyzed in this paper in $(\ell,b)$ space ({\it Top}) and (Age, Distance) space ({\it Bottom}).
The age-distance distributions are predictions from the {\sl Galaxia} model
for stars satisfying the RAVE selection criteria.
\label{fig:rave_stat1}}
\end{figure}
\begin{figure}
\centering \includegraphics[width=0.50\textwidth,trim=10mm 50mm 0mm 0mm,clip=true]{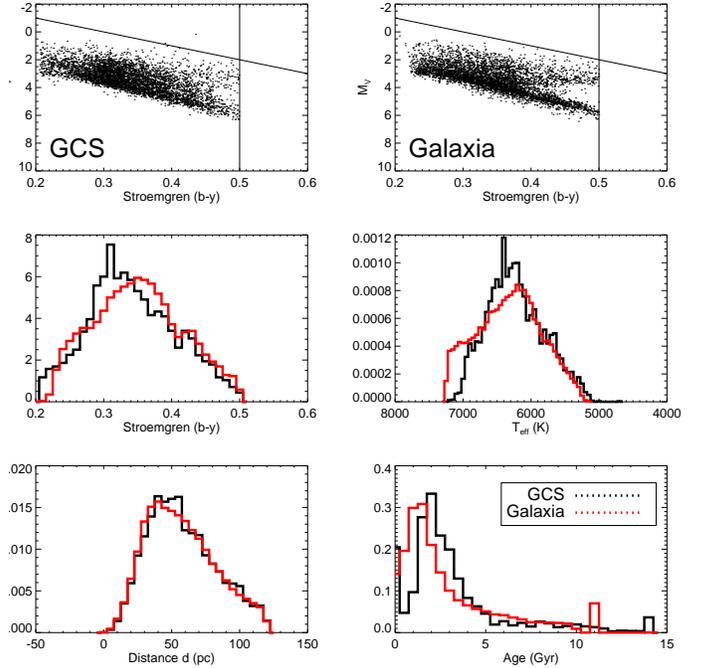}\caption{Distribution of GCS stars as a function of color, temperature, distance and age. Shown alongside are
results of a mock sample created using {\sl Galaxia}
but without observational uncertainties.
The top panel shows the distribution in the $(b-y,M_V)$ plane; the colors
span the range $0.205<(b-y)<0.5$.
The magnitude limits are a function of color and are taken from
\citet{2004A&A...418..989N}.
The line represents the equation $M_V=10(b-y)-3$ and is used
to mimic the selective avoidance of giants in GCS.
A selection of $d<0.12\kpc$ and $T_{\rm eff}>7244$~K is also applied.
The temperature and ages (maximum likelihood Padova)
are from \citet{2011A&A...530A.138C}.
\label{fig:gcs2}}
\end{figure}

\subsection{GCS survey}
We fit the models to all six phase-space coordinates
of a subset of the  16682 F and G type
main-sequence stars in the GCS \citep{2004A&A...418..989N,2009A&A...501..941H}.
A mock  GCS sample was extracted from the model as in \citet{2011ApJ...730....3S}.
Velocities and temperatures are available for 13382 GCS stars.
We found that while {\sl Galaxia} predicts less than one halo star
in the GCS sample for a distance less than $120$ pc,
when plotted in $({\rm [Fe/H]},v_\phi)$ plane, the GCS has
29 stars with ${\rm [Fe/H]}<-1.2$ and highly
negative values of $v_\phi$ (as expected for halo stars).
Following \citet{2010MNRAS.403.1829S},
we identify these as halo stars and exclude them from our analysis.

The GCS catalog is complete for F and G type stars within a volume
given by $r < 40$ pc and $V \sim 8$ in magnitude;
within these limits there are only 1342 stars. But since GCS is a
color-magnitude limited survey, there is no need to restrict
the analysis to a volume complete sample.  In
\citet{2004A&A...418..989N}
magnitude completeness as a function of color is provided
and we use this (their \S 2.2). There is some ambiguity about the
coolest dwarfs which were added for declination $\delta<-26^{\circ}$;
from information gleaned from \citet{2004A&A...418..989N},
we could not find a suitable way to take this into account.

We also applied some additional restrictions on the sample.
For example, we restrict our analysis to stars with distance less than 120 pc,
so as to avoid stars with large distance errors.
The GCS survey selectively avoids giants. To mimic this we
use the following selection function $M_V<10(b-y)-3$.
The predicted temperature distributions
show a mismatch with models, in particular, there are too many hot stars. Using
\citet{2011A&A...530A.138C} temperatures, which are more accurate,
we found an upper limit
on $T_{\rm eff}$ of 7244~K, which was applied to the models.

After the above mentioned cuts, the final sample consisted
of 5201 stars. Note, we do not remove possible binary stars
as this will further reduce the number of stars. In future,
we think it will be instructive to check if there is any
systematic associated with the inclusion or exclusion
of binaries.
The black histograms in \fig{gcs2} show the distribution of these stars, while
the red histograms show the predictions of the model. At the hot end, the
temperature distributions of model and data are  still discrepant, but the
distance distributions agree nicely.
The model's age distribution is
qualitatively correct but differences can also be seen.
The plotted GCS ages are maximum likelihood Padova ages and
there can be systematics associated with this.
A more quantitative
comparison would require estimating the  ages of model stars
in the same way as done by GCS and taking into account
uncertainties and systematics which we do not do here.
The peak in the model at 11 Gyr is due to the
thick disc having a fixed age. The peaks in the data at 0 and $14\Gyr$ are
most likely due to caps employed while estimating ages. 
The color
distribution in GCS shows a peak at around $b-y=0.3$, which could be due to
an unknown selection effect.  The bump at $b-y \sim 0.43$, which is also seen
in models, is due to turnoff stars.  Overall, we think our modelling
reproduces to a good degree the selection function of the GCS stars.

\section{Model Fitting techniques}\label{sec:modtech}
If $y_i$ are the observed properties of a star, we can
describe the observed data by $y=\left\{y_i \in \mathbb{R}^d, 0<i<N\right\}$.
Also, let $\theta$ be the set of parameters that define the model.
Our job is to compute
\be
p(\theta|y) \propto p(y|\theta)p(\theta)
\ee
where $p(y|\theta)=\prod_i p(y_i|\theta)$. We employ an MCMC scheme to
estimate $p(\theta|y)$ and assume
a uniform prior on $\theta$. We now discuss how to compute
$p(y_i|\theta)$.

Generally, a model of a galaxy gives the probability density $p({\bf
r,v},\tau,Z,m|\theta)$.
For RAVE, the observed quantities are $v_{\rm los},l$ and $b$,
while for GCS they are $l,b,r,v_l,v_b$ and $v_{\rm los}$. Since
quantities like $\tau,Z$ and $m$ are unknown,
one has to compute the marginal probability density by integration.
For RAVE, the required marginal density is
{\small
\be
p(\ell,b,v_{\rm los}|\theta) &= & \int p(\ell,b,r,\tau,Z,m,v_l,v_b,v_{\rm
los}|\theta)   \nonumber \\
& & \times S(\ell,b,\tau,Z,m)\: dr\: d\tau\: dZ\: dm\:   dv_l\: dv_b,
\ee
}
and for GCS it is
\be
p(\ell,b,r,v_l,v_b,v_{\rm los}|\theta) &= & \int
p(\ell,b,r,\tau,Z,m,v_l,v_b,v_{\rm los}|\theta)   \nonumber \\
& & \times S(\ell,b,\tau,Z,m)\: d\tau\: dZ\: dm .
\ee
Here $S(\ell,b,\tau,Z,m)$ is the selection function specifying how the stars
were preselected in the data. The actual selection is on photometric
magnitude which in turn is a function of  $\tau,Z$ and $m$.

For the kinds of models explored here, the computations are considerably
simplified due to the fact that
\be
p(\ell,b,r,\tau,Z,m,v_l,v_b,v_{\rm los}|\theta) & = &
p(v_l,v_b,v_{\rm los}|\ell,b,r,\tau,\theta)  \nonumber \\
\times p(\ell,b,r,\tau,Z,m|\theta_S) ,
\label{equ:prv_theta}
\ee
for which $\theta_S$ is the set of model parameters that govern the spatial
distribution of stars and $\theta$ is the set of model parameters that govern
the kinematic distribution of stars.
The term
$p(\ell,b,r,\tau,Z,m|\theta_S)$ is invariant in our analysis, and this is the
main assumption that we make. In other words we assume
star formation rate (SFR), initial mass function (IMF),
scale length of disc, age scale-height relation, age metallicity relation and
radial metallicity gradient for the disc.  All
these distributions can be constrained by the stellar photometry.  The
distribution $p(v_l,v_b,v_{\rm los}|\ell,b,r,\tau,\theta)$ represents the
kinematics, which is what we explore.  It should be noted that the model
$p(\ell,b,r,\tau,Z,m|\theta_S)$ that we use has been shown to satisfy the
number count of stars \citep{2003A&A...409..523R,2011ApJ...730....3S}.
In a fully self consistent model, the scale height,
the
vertical stellar velocity dispersion and the potential would 
all be related to each other and this is something we would like to address in
future.

We can now integrate the last term in \equ{prv_theta} over $m$ and $Z$ such that
\be
p(\ell,b,r,v_l,v_b,v_{\rm los},\tau|\theta) &= &  p(v_l,v_b,v_{\rm
los}|\ell,b,r,\tau,\theta)   \nonumber \\
& & \times p(\ell,b,r,\tau| \theta_S,S)
\ee
where
\be
p(\ell,b,r,\tau| \theta_S,S) & =& \int \int p(\ell,b,r,\tau,Z,m|\theta_S)   \nonumber\\
& & \times S(\ell,b,\tau,Z,m)\: dZ\: dm.
\ee
The term $p(\ell,b,r,\tau| \theta_S,S)$ is computed numerically
using the code
{\sl {\sl Galaxia}} \citep{2011ApJ...730....3S}.
Galaxia, uses isochrones from the Padova database
to compute photometric magnitudes for the model stars
\citep{2008A&A...482..883M,1994AAS..106..275B}.
We first generate a fiducial set
of stars satisfying the color-magnitude range of the survey.
Then we apply the selection function and reject stars that
do not satisfy the constraints of the survey. The accepted stars
are then binned in $(\ell,b,r,\tau)$ space.
Since, the GCS  is local to the Sun,
we use the following approximation $p(\ell,b,r,\tau| \theta_S,S) \propto p(\tau| \theta_S,S)$. The probability distribution
in $(\ell,b,r,\tau)$ space for RAVE is shown in \fig{rave_stat1}.

For RAVE, we have to integrate over four variables ($r,\tau,v_l,v_b$),
but for GCS we integrate over only $\tau$.
The 4D marginalization for RAVE poses a serious
computational challenge for data as large as the RAVE survey.
For Gaussian distribution functions, the integral over $v_l$ and
$v_b$ can be performed analytically to give an analytic expression for
$p(v_{\rm los}|\ell,b,r,\tau,Z,\theta)$, but in general it cannot
be done analytically. Hence, we try two new methods. The first method
is fast but has inflated uncertainties.
The second method is slower to converge
but gives correct estimates of uncertainties.
Given these strengths
and limitations, we use a combined strategy that makes
best use of both the methods.

We use the first `sampling and projection' method to get an initial estimate of $\theta$
and also its covariance matrix. These are then used in the second `data augmentation'
method.  The initial estimate reduces the `burn in'
time, while the covariance matrix eliminates the need to
tune the widths of the proposal distributions.
In general we use an adaptive MCMC scheme, which avoids manual
tuning of the widths of the proposal distributions \citep{andrieu2008tutorial}.
At regular intervals,
we compute the covariance matrix and scale it so as to achieve
the desired acceptance ratio for the given number of parameters
\citet{gelman1996efficient}. We now discuss
the two methods in more detail.

\subsection{MCMC using sampling and projection} Instead of doing the
computationally intensive marginalization, at each step of the Markov chain
of model parameters, we generate a sample of stars by Monte-Carlo sampling
the current model subject to the selection function.  Binning these stars in
$(\ell,b,v_{\rm los})$ space then gives an estimate of $p(\ell,b,v_{\rm
los}|\theta)$.  Note that, given the
stochastic nature of our  estimate of $p(\ell,b,v_{\rm
los}|\theta)$, the standard
Metropolis-Hastings algorithm had to be altered to avoid the simulation from
getting stuck at a stochastic maximum of the likelihood.

\subsection{MCMC using data augmentation}
Instead of marginalizing one can treat the nuisance parameters
as unknown parameters and estimate them alongside other parameters.
This constitutes what is known as a sampling based approach for computing the
marginal densities. The basic form of this scheme was introduced by
\citet{tanner1987calculation} and later extended by
\citet{gelfand1990sampling}.
Let  $x=\left\{x_i \in \mathbb{R}^d
,0<i<N \right\}$ be an extra set of variables that are needed
by the model to compute the probability density.
Then we can write
\be
p(\theta,x| y)\propto p(x,y|\theta) p(\theta).
\ee
where $p(x,y|\theta)=\prod_{i} p(x_i,y_i|\theta)$, and
$p(x_i,y_i|\theta)$ is a function which is known and relatively
easy to compute.
For example, for the RAVE data $y_i=\left\{l_i,b_i,v_{i,{\rm
los}}\right\}$ and  $x=\left\{r_i,\tau_i,v_{l,i},v_{b,i}\right\}$.
Due to the unusually large number of parameters, it is difficult to get
satisfactory acceptance rates with the standard Metropolis-Hastings
scheme without making the widths of the
proposal distributions extremely small. Thus the chains
would take an unusually long time to mix. To solve this, one uses
the Metropolis scheme with Gibbs sampling (MWG)
\citep{tierney1994markov}. The MWG scheme is also useful
for solving hierarchical Bayesian models, and its
application for 3D extinction mapping is discussed in
\citet{2012MNRAS.427.2119S}.
In our case, the Gibbs step consists of first sampling $x$
from the conditional density $p(x|y,\theta)$ and then $\theta$
from the conditional density $p(\theta|y,x)$.
The sampling in each Gibbs step is done using the
Metropolis-Hastings algorithm.

\subsection{Goodness of fit}
To assess the ability of a model to fit the data, we compute
an approximate  reduced $\chi^2$ value.
To accomplish this, first we bin the data in the 
observational space.
For RAVE, we bin the data in $(l,b,v_{\rm los})$
space with bins of size 859 deg$^2$ and $5\kms$.
Angular binning was done using the HEALPIX scheme.
For GCS, we
bin the  $U,V$ and $W$ components of velocity separately
with bins of size $5\kms$.
Next, an N-body realization of a given model was created
satisfying the same constraints as the data,
The reduced $\chi^2$ between the data and the model was 
then computed as
\be
\chi_{\rm red}^2=\Bigl\langle \sum_i \frac{(n_i-m_i/f_{\rm
sample})^2}{n_i+m_i/f_{\rm sample}^2} \Bigr\rangle {\ \rm
for\ } n_i>0.
\ee
Here,  $n_i$ is the
number of data points in a bin, $m_i$ the number of
model points in the same bin and
$f_{\rm sample}=\sum_i m_i/\sum_i n_i$ is the sampling fraction.
Choosing, $f_{\rm sample}$ to be very high one can increase
the precision of the estimate, but then it increases the
computational cost. For RAVE $f_{\rm sample}$ was 1
while for GCS it was 10.
To decrease the stochasticity in the estimate, we computed
the mean over 30 random estimates
$\langle \chi_{\rm red}^2 \rangle=\sum_{k=1}^{30} \chi_{\rm red,k}^2/30$.

The reduced $\chi^2$ as computed above, has its
limitations. Firstly, it is not an accurate estimator
of the goodness of fit. Secondly, $\chi^2$ value is sensitive to the
choice of bin size and $f_{\rm
sample}$. Hence, it is not advisable to estimate statistical significance
using our reduced $\chi^2$. However, the reduced $\chi^2$ should
be good enough to qualitatively compare the goodness of
fit of two models.

\subsection{Tests using synthetic data}
We now describe tests in which mock data are sampled from the distribution
function and then fitted using the MCMC machinery.  These tests serve two
main purposes. First, they determine if our MCMC scheme works correctly.
Secondly, they tell us which parameters can be recovered and with what
accuracy.  We study two classes of models based on (1) the Gaussian DF and
(2) the Shu DF.  Additionally, we study two types of mock data, one
corresponding to the RAVE survey and the other to the GCS survey.  For GCS we
also study models where $\Theta_0$ is fixed.  Altogether this leads to 6
different types of tests.

The results of these tests are summarized in Tables \ref{tab:tbmock1} and
\ref{tab:tbmock2}. The difference of a parameter $p$ from input
values divided by uncertainty $\sigma_p$ measures the confidence
of recovering the parameter. To aid the comparison, we
color the values if they differ significantly from the input values: $|\delta p|/\sigma_p<2$
(black), $2<|\delta p|/\sigma_p<3$ (blue).
It can be seen that all parameters are recovered
within the 3$\sigma$ range as given by the error bars.
Ideally to check the systematics, the fitting should be
repeated multiple times and the mean values should be
compared with  input values. However, the MCMC simulations
being computationally very expensive we report results
with only one independent data sample for each of the test cases.

It can be seen that GCS type data cannot properly constrain $\Theta_0$.  This
is because the GCS sample is very local to the Sun.  Keeping $\Theta_0$ free
also has the undesirable effect of increasing the uncertainty of
$R_\sigma^{\rm thin}$ and $R_\sigma^{\rm thick}$.  For Gaussian models, it is
easy to see from \equ{stromberg} that the effect of changing $\Theta_0$ can
be compensated by a change in $R_\sigma^{\rm thin}$ and $R_\sigma^{\rm
thick}$.  Given these limitations, when analyzing GCS we keep $\Theta_0$
fixed to 226.87 km s$^{-1}$, a value that was used by
\citet{2011ApJ...730....3S} in the {\sl Galaxia} code.

The Solar motion is constrained well by
both surveys, but better by RAVE. RAVE is also clearly better in
constraining thick-disc parameters than GCS, mainly because
the GCS has very few thick-disc stars ({\sl Galaxia} estimates
it to be 6\% of the overall GCS sample).
{\it Across all parameters,
for Shu models $\beta_z$ is the only parameter which is constrained better by
GCS than by RAVE}. This is because RAVE only has radial
velocities. This means that only those stars that lie towards the  pole
can carry meaningful information about the vertical motion, and such
stars constitute a much smaller subset of the whole RAVE sample.
This suggests that one can use
the $\beta_z$ value from GCS when fitting the RAVE data, as we show below.

\begin{table}
\caption{\label{tab:tbmock1}
Tests on mock data: Constraints on model parameters with Gaussian
distribution function. The model runs are named as follows; survey
name as RAVE or GCS, type of model as GAU for Gaussian and SHU for Shu.
Parameters that do not have error bars were fixed.
Velocities are in $\kms$ and distances in ${\rm kpc}$}
\centering
\begin{tabular}{|l|l|l|l|l|} \hline
Model & GCS GAU & GCS GAU & RAVE GAU & Input \\ \hline
$U_{\odot}$ &  {$11.12_{-0.41}^{+0.43}$} &  {$11.17_{-0.39}^{+0.39}$} &  {$11.22_{-0.16}^{+0.15}$} & $11.1$ \\ \hline
$V_{\odot}$ &  {$5.8_{-1.9}^{+1.8}$} &  {$8.6_{-1.3}^{+1.3}$} & \textcolor{blue} {$8.16_{-0.24}^{+0.29}$} & $7.5$ \\ \hline
$W_{\odot}$ &  {$7.14_{-0.19}^{+0.19}$} &  {$7.35_{-0.18}^{+0.19}$} &  {$7.377_{-0.087}^{+0.092}$} & $7.25$ \\ \hline
$\sigma_R^{\rm thin}$ &  {$38.5_{-1.6}^{+1.7}$} &  {$42.7_{-1.6}^{+1.6}$} &  {$40.45_{-0.84}^{+0.56}$} & $40$ \\ \hline
$\sigma_{\phi}^{\rm thin}$ &  {$28.8_{-1}^{+1.1}$} &  {$28.4_{-1.1}^{+1.1}$} &  {$27.7_{-0.5}^{+0.42}$} & $28.3$ \\ \hline
$\sigma_z^{\rm thin}$ &  {$25.03_{-0.84}^{+0.86}$} &  {$25.89_{-0.86}^{+0.87}$} &  {$25.09_{-0.72}^{+0.6}$} & $25$ \\ \hline
$\sigma_R^{\rm thick}$ &  {$63.3_{-3.8}^{+3.8}$} &  {$55.3_{-4}^{+4.2}$} &  {$60.62_{-0.68}^{+0.55}$} & $60$ \\ \hline
$\sigma_{\phi}^{\rm thick}$ &  {$47.8_{-2.9}^{+3.1}$} &  {$43.7_{-3.1}^{+3.2}$} &  {$42.02_{-0.4}^{+0.45}$} & $42.4$ \\ \hline
$\sigma_z^{\rm thick}$ &  {$34.1_{-2.1}^{+2.3}$} &  {$32.8_{-2.3}^{+2.3}$} &  {$35.19_{-0.52}^{+0.58}$} & $35$ \\ \hline
$\beta_R$ &  {$0.183_{-0.025}^{+0.025}$} & \textcolor{blue} {$0.249_{-0.023}^{+0.021}$} &  {$0.2079_{-0.015}^{+0.0094}$} & $0.2$ \\ \hline
$\beta_{\phi}$ &  {$0.216_{-0.022}^{+0.023}$} &  {$0.197_{-0.023}^{+0.022}$} &  {$0.177_{-0.016}^{+0.013}$} & $0.2$ \\ \hline
$\beta_z$ &  {$0.38_{-0.022}^{+0.022}$} &  {$0.401_{-0.022}^{+0.02}$} &  {$0.368_{-0.03}^{+0.025}$} & $0.37$ \\ \hline
$1/R_{\sigma}^{\rm thin}$ & $0.145_{-0.064}^{+0.067}$ & $0.055_{-0.061}^{+0.077}$ & $0.072_{-0.0058}^{+0.005}$ & $0.072$ \\ \hline
$1/R_{\sigma}^{\rm thick}$ & $0.107_{-0.034}^{+0.04}$ & $0.133_{-0.074}^{+0.065}$ & $0.1341_{-0.0029}^{+0.0029}$ & $0.132$ \\ \hline
$\Theta_0$ & \textcolor{magenta}{$233$} &  {$265_{-60}^{+63}$} &  {$236_{-1.4}^{+1.7}$} & $233$ \\ \hline
$R_0$ & \textcolor{magenta}{$8$} & \textcolor{magenta}{$8$} & \textcolor{magenta}{$8$} & $8$ \\ \hline
$\alpha_{z}$ & \textcolor{magenta}{$0.047$} & \textcolor{magenta}{$0.047$} & \textcolor{blue} {$0.0432_{-0.0019}^{+0.0015}$} & $0.047$ \\ \hline
$\alpha_{R}$ & \textcolor{magenta}{$0$} & \textcolor{magenta}{$0$} & \textcolor{magenta}{$0$} & $0$ \\ \hline
$\chi_{\rm red}^2$ & 1.09 & 1.00 & 0.935 \\ \hline
\end{tabular}
\caption{\label{tab:tbmock2}
Tests on mock data: Constraints on model parameters with Shu distribution function}
\begin{tabular}{|l|l|l|l|l|} \hline
Model & GCS SHU & GCS SHU & RAVE SHU & Input \\ \hline
$U_{\odot}$ &  {$11.28_{-0.41}^{+0.42}$} &  {$11.16_{-0.41}^{+0.42}$} &  {$11.27_{-0.14}^{+0.12}$} & $11.1$ \\ \hline
$V_{\odot}$ &  {$7.14_{-0.36}^{+0.34}$} &  {$7.35_{-0.67}^{+0.79}$} & \textcolor{blue} {$7.94_{-0.15}^{+0.17}$} & $7.5$ \\ \hline
$W_{\odot}$ &  {$6.95_{-0.2}^{+0.19}$} &  {$6.99_{-0.2}^{+0.2}$} &  {$7.26_{-0.088}^{+0.079}$} & $7.25$ \\ \hline
$\sigma_R^{\rm thin}$ &  {$41_{-1.1}^{+1.1}$} &  {$40.7_{-1.2}^{+1.1}$} & \textcolor{blue} {$41.19_{-0.6}^{+0.47}$} & $40$ \\ \hline
$\sigma_z^{\rm thin}$ &  {$25.18_{-0.84}^{+0.84}$} &  {$24.9_{-0.92}^{+0.95}$} &  {$24.62_{-0.65}^{+0.81}$} & $25$ \\ \hline
$\sigma_R^{\rm thick}$ &  {$45.2_{-3.5}^{+3.6}$} &  {$44.3_{-4}^{+3.9}$} &  {$46.1_{-0.58}^{+0.61}$} & $45$ \\ \hline
$\sigma_z^{\rm thick}$ &  {$36.8_{-2.4}^{+2.6}$} &  {$32.3_{-2.5}^{+2.4}$} &  {$34.3_{-0.51}^{+0.52}$} & $35$ \\ \hline
$\beta_R$ &  {$0.203_{-0.016}^{+0.016}$} &  {$0.201_{-0.017}^{+0.017}$} &  {$0.211_{-0.013}^{+0.01}$} & $0.2$ \\ \hline
$\beta_z$ &  {$0.379_{-0.021}^{+0.021}$} &  {$0.371_{-0.024}^{+0.023}$} &  {$0.331_{-0.025}^{+0.036}$} & $0.37$ \\ \hline
$1/R_{\sigma}^{\rm thin}$ & $0.0696_{-0.0075}^{+0.0071}$ & $0.074_{-0.011}^{+0.015}$ & $0.0682_{-0.0026}^{+0.0027}$ & $0.072$ \\ \hline
$1/R_{\sigma}^{\rm thick}$ & $0.133_{-0.016}^{+0.016}$ & $0.131_{-0.017}^{+0.018}$ & $0.1307_{-0.0027}^{+0.0025}$ & $0.132$ \\ \hline
$\Theta_0$ & \textcolor{magenta}{$233$} &  {$224_{-20}^{+33}$} &  {$235.1_{-1.3}^{+1.3}$} & $233$ \\ \hline
$R_0$ & \textcolor{magenta}{$8$} & \textcolor{magenta}{$8$} & \textcolor{magenta}{$8$} & $8$ \\ \hline
$\alpha_{z}$ & \textcolor{magenta}{$0.047$} & \textcolor{magenta}{$0.047$} & \textcolor{blue} {$0.0427_{-0.0018}^{+0.0019}$} & $0.047$ \\ \hline
$\alpha_{R}$ & \textcolor{magenta}{$0$} & \textcolor{magenta}{$0$} & \textcolor{magenta}{$0$} & $0$ \\ \hline
$\chi_{\rm red}^2$ & 0.960 & 0.996 & 0.928 \\ \hline
\end{tabular}
\end{table}

\begin{table}
\caption{\label{tab:tbfid}
Fiducial model parameters: Velocities are in $\kms$ and distances in
${\rm kpc}$}
\centering
\begin{tabular}{|l|l|l|} \hline
Model & {\sl Galaxia} & Equivalent Besan\c{c}on \\ \hline
$U_{\odot}$  & $11.1$ & $10.3$ \\ \hline
$V_{\odot}$ & $12.24$ & $6.3$ \\ \hline
$W_{\odot}$ & $7.25$ & $5.9$ \\ \hline
$\sigma_R^{\rm thin}$  & $50$ & $50$ \\ \hline
$\sigma_{\phi}^{\rm thin}$  & $32.3$ & $32.3$ \\ \hline
$\sigma_z^{\rm thin}$  & $21$  & $21$ \\ \hline
$\sigma_R^{\rm thick}$  & $67$ & $67$ \\ \hline
$\sigma_{\phi}^{\rm thick}$ & $51$ & $51$ \\ \hline
$\sigma_z^{\rm thick}$  & $42$ & $42$ \\ \hline
$\beta_R$  & $0.33$ & $0.33$ \\ \hline
$\beta_{\phi}$  & $0.33$ & $0.33$ \\ \hline
$\beta_z$   & $0.33$ & $0.33$ \\ \hline
$\tau_{\rm sat}$  & 6.5 Gyr  & 6.5 Gyr \\ \hline
$1/R_{\sigma}^{\rm thin}$  & $0.133$ & $0.096 (0.114)$ \\ \hline
$1/R_{\sigma}^{\rm thick}$  & $0.133$ & $0.176 (0.2)$ \\ \hline
$R_{0}$  & $8.0$ & $8.5 (8.0)$ \\ \hline
$\Theta_0$  & $226.84$& $220.0$\\ \hline
$R_d$  & $2.5$ & $2.5$\\ \hline
\end{tabular}
\end{table}
\begin{table*}
\caption{\label{tab:tbgauss}
Constraints on model parameters with the Gaussian distribution
function. Parameters that do not have error bars were fixed. Missing
values imply parameters that are not applicable for that model.
The model runs are named as follows; survey
name as RAVE or GCS, type of model as GAU for Gaussian and SHU for Shu.
Velocities are in $\kms$ and distances in ${\rm
kpc}$. Quoted uncertainties  are purely
random and do not include systematics.}
\centering
\begin{tabular}{|l|l|l|l|l|l|l|} \hline
Model & GCS GAU & GCS GAU & GCS GAU & RAVE GAU & RAVE GAU & RAVE GAU \\ \hline
$U_{\odot}$ & $10.16_{-0.42}^{+0.41}$ & $10.28_{-0.43}^{+0.43}$ & $10.34_{-0.42}^{+0.42}$ & $11.66_{-0.15}^{+0.16}$ & $11.45_{-0.14}^{+0.14}$ & $11.25_{-0.15}^{+0.15}$ \\ \hline
$V_{\odot}$ & $6.6_{-1.4}^{+1.3}$ & $6.33_{-0.97}^{+0.93}$ & $9.68_{-0.26}^{+0.26}$ & $15.01_{-0.42}^{+0.37}$ & $8_{-0.28}^{+0.3}$ & $7.38_{-0.12}^{+0.1}$ \\ \hline
$W_{\odot}$ & $7.14_{-0.18}^{+0.19}$ & $7.11_{-0.19}^{+0.19}$ & $7.14_{-0.18}^{+0.18}$ & $7.692_{-0.082}^{+0.099}$ & $7.688_{-0.091}^{+0.085}$ & $7.625_{-0.082}^{+0.088}$ \\ \hline
$\sigma_R^{\rm thin}$ & $41.2_{-1.3}^{+1.4}$ & $47_{-1.1}^{+1.1}$ & $41.5_{-1.3}^{+1.4}$ & $36.6_{-1.1}^{+1}$ & $39.26_{-0.69}^{+0.67}$ & $39.69_{-0.65}^{+0.62}$ \\ \hline
$\sigma_{\phi}^{\rm thin}$ & $27.12_{-0.86}^{+0.89}$ & $31.61_{-0.79}^{+0.8}$ & $27.83_{-0.88}^{+0.88}$ & $24.97_{-0.36}^{+0.43}$ & $25.56_{-0.37}^{+0.33}$ & $25.34_{-0.33}^{+0.35}$ \\ \hline
$\sigma_z^{\rm thin}$ & $23.74_{-0.74}^{+0.79}$ & $27.28_{-0.63}^{+0.64}$ & $23.89_{-0.74}^{+0.79}$ & $24.22_{-0.47}^{+0.64}$ & $25.69_{-0.2}^{+0.22}$ & $25.92_{-0.2}^{+0.21}$ \\ \hline
$\sigma_R^{\rm thick}$ & $65.9_{-3.7}^{+4.1}$ &  & $67.7_{-2.7}^{+2.7}$ & $58.74_{-0.79}^{+0.91}$ & $58.43_{-0.76}^{+0.86}$ & $57.87_{-0.56}^{+0.58}$ \\ \hline
$\sigma_{\phi}^{\rm thick}$ & $40.9_{-3.1}^{+3.3}$ &  & $40_{-2.8}^{+2.9}$ & $40.47_{-0.48}^{+0.51}$ & $37.16_{-0.53}^{+0.5}$ & $38.37_{-0.54}^{+0.48}$ \\ \hline
$\sigma_z^{\rm thick}$ & $38.5_{-2.5}^{+2.8}$ &  & $38.7_{-2.6}^{+2.7}$ & $40.55_{-0.49}^{+0.46}$ & $40.4_{-0.5}^{+0.5}$ & $39.41_{-0.48}^{+0.48}$ \\ \hline
$\beta_R$ & $0.201_{-0.019}^{+0.019}$ & $0.268_{-0.014}^{+0.015}$ & $0.204_{-0.019}^{+0.019}$ & $0.06_{-0.029}^{+0.023}$ & $0.135_{-0.015}^{+0.015}$ & $0.164_{-0.013}^{+0.012}$ \\ \hline
$\beta_{\phi}$ & $0.271_{-0.019}^{+0.019}$ & $0.349_{-0.015}^{+0.016}$ & $0.284_{-0.019}^{+0.019}$ & $0.132_{-0.013}^{+0.014}$ & $0.17_{-0.012}^{+0.012}$ & $0.164_{-0.012}^{+0.012}$ \\ \hline
$\beta_z$ & $0.36_{-0.021}^{+0.02}$ & $0.432_{-0.016}^{+0.015}$ & $0.365_{-0.02}^{+0.02}$ & $0.312_{-0.02}^{+0.026}$ & \textcolor{magenta}{$0.37$} & \textcolor{magenta}{$0.37$} \\ \hline
$1/R_{\sigma}^{\rm thin}$ & $0.171_{-0.043}^{+0.046}$ & $0.179_{-0.027}^{+0.028}$ & \textcolor{magenta}{$0.073$} & $-0.0556_{-0.0077}^{+0.0078}$ & $0.0188_{-0.0053}^{+0.0055}$ & \textcolor{magenta}{$0.073$} \\ \hline
$1/R_{\sigma}^{\rm thick}$ & $0.148_{-0.035}^{+0.04}$ &  & \textcolor{magenta}{$0.132$} & $0.1123_{-0.0043}^{+0.0044}$ & $0.0907_{-0.0036}^{+0.0035}$ & \textcolor{magenta}{$0.132$} \\ \hline
$\Theta_0$ & \textcolor{magenta}{$226.84$} & \textcolor{magenta}{$226.84$} & \textcolor{magenta}{$233$} & $207.2_{-1.9}^{+1.9}$ & $229.2_{-2}^{+1.8}$ & $234.1_{-1.4}^{+1.4}$ \\ \hline
$R_0$ & \textcolor{magenta}{$8$} & \textcolor{magenta}{$8$} & \textcolor{magenta}{$8$} & \textcolor{magenta}{$8$} & \textcolor{magenta}{$8$} & \textcolor{magenta}{$8$} \\ \hline
$\alpha_{z}$ & \textcolor{magenta}{$0$} & \textcolor{magenta}{$0$} & \textcolor{magenta}{$0.047$} & \textcolor{magenta}{$0$} & $0.0738_{-0.0023}^{+0.0021}$ & \textcolor{magenta}{$0.047$} \\ \hline
$\alpha_{R}$ & \textcolor{magenta}{$0$} & \textcolor{magenta}{$0$} & \textcolor{magenta}{$0$} & \textcolor{magenta}{$0$} & \textcolor{magenta}{$0$} & \textcolor{magenta}{$0$} \\ \hline
$\chi_{\rm red}^2$ RAVE & 2.55 & 3.19 & 2.49 & 1.89 & 1.64 & 1.79 \\ \hline
$\chi_{\rm red}^2$ GCS & 3.09 & 3.48 & 3.15 & 6.60 & 5.81 & 5.10 \\ \hline
\end{tabular}
\caption{\label{tab:tbshu}
Constraints on model parameters with the Shu distribution function. See
\tab{tbgauss} for further description.}
\begin{tabular}{|l|l|l|l|l|l|l|l|} \hline
Model & GCS SHU & GCS SHU & GCS SHU & RAVE SHU & RAVE SHU & RAVE SHU & RAVE SHU \\ \hline
$U_{\odot}$ & $10.02_{-0.4}^{+0.39}$ &
$10.16_{-0.4}^{+0.39}$ & $10.23_{-0.4}^{+0.39}$ &
$11.2_{-0.13}^{+0.13}$ & $10.92_{-0.14}^{+0.13}$ & $10.96_{-0.13}^{+0.14}$ & $11.05_{-0.16}^{+0.15}$ \\ \hline
$V_{\odot}$ & $9.95_{-0.3}^{+0.3}$ & $9.81_{-0.28}^{+0.28}$ & $9.83_{-0.29}^{+0.3}$ & $9.71_{-0.11}^{+0.12}$ & $7.53_{-0.16}^{+0.16}$ & $7.53_{-0.16}^{+0.16}$ & $7.62_{-0.16}^{+0.13}$ \\ \hline
$W_{\odot}$ & $7.14_{-0.19}^{+0.19}$ & $7.13_{-0.19}^{+0.18}$ & $7.12_{-0.19}^{+0.18}$ & $7.536_{-0.086}^{+0.085}$ & $7.542_{-0.093}^{+0.089}$ & $7.539_{-0.09}^{+0.095}$ & $7.553_{-0.09}^{+0.086}$ \\ \hline
$\sigma_R^{\rm thin}$ & $38.14_{-0.94}^{+0.98}$ & $39.99_{-0.91}^{+0.91}$ & $42.71_{-0.8}^{+0.83}$ & $42.37_{-0.66}^{+0.61}$ & $39.78_{-0.73}^{+0.81}$ & $39.67_{-0.72}^{+0.63}$ & $39.56_{-0.7}^{+0.66}$ \\ \hline
$\sigma_z^{\rm thin}$ & $23.39_{-0.73}^{+0.77}$ & $23.63_{-0.8}^{+0.85}$ & $25.91_{-0.6}^{+0.64}$ & $26.85_{-0.92}^{+0.85}$ & $24.7_{-0.66}^{+0.66}$ & $25.73_{-0.21}^{+0.21}$ & $25.72_{-0.25}^{+0.23}$ \\ \hline
$\sigma_R^{\rm thick}$ & $70.1_{-5.5}^{+3.7}$ & $45.9_{-1.8}^{+1.8}$ &  & $38.84_{-0.96}^{+1.2}$ & $42.31_{-0.9}^{+1}$ & $42.43_{-1}^{+0.95}$ & $43.23_{-1.1}^{+0.96}$ \\ \hline
$\sigma_z^{\rm thick}$ & $39_{-3.3}^{+3.1}$ & $32.6_{-2.2}^{+2.3}$ &  & $29.15_{-0.79}^{+0.87}$ & $34.66_{-0.58}^{+0.61}$ & $34.3_{-0.57}^{+0.51}$ & $34.48_{-0.53}^{+0.54}$ \\ \hline
$\beta_R$ & $0.213_{-0.014}^{+0.014}$ & $0.237_{-0.013}^{+0.013}$ & $0.273_{-0.011}^{+0.011}$ & $0.236_{-0.011}^{+0.011}$ & $0.198_{-0.014}^{+0.014}$ & $0.195_{-0.013}^{+0.011}$ & $0.192_{-0.013}^{+0.012}$ \\ \hline
$\beta_z$ & $0.361_{-0.02}^{+0.02}$ & $0.366_{-0.021}^{+0.021}$ & $0.415_{-0.016}^{+0.016}$ & $0.398_{-0.029}^{+0.03}$ & $0.328_{-0.024}^{+0.027}$ & \textcolor{magenta}{$0.37$} & \textcolor{magenta}{$0.37$} \\ \hline
$1/R_{\sigma}^{\rm thin}$ & $0.0665_{-0.0086}^{+0.0084}$ & \textcolor{magenta}{$0.073$} & $0.0771_{-0.0061}^{+0.0059}$ & $0.0673_{-0.0028}^{+0.0028}$ & $0.0722_{-0.0032}^{+0.0035}$ & $0.073_{-0.003}^{+0.0037}$ & $0.0724_{-0.0031}^{+0.0031}$ \\ \hline
$1/R_{\sigma}^{\rm thick}$ & $0.0086_{-0.0066}^{+0.022}$ & \textcolor{magenta}{$0.132$} &  & $0.1555_{-0.0064}^{+0.0046}$ & $0.1335_{-0.0056}^{+0.0046}$ & $0.1328_{-0.0051}^{+0.005}$ & $0.13_{-0.0046}^{+0.0056}$ \\ \hline
$\Theta_0$ & \textcolor{magenta}{$226.84$} & \textcolor{magenta}{$232$} & \textcolor{magenta}{$226.84$} & $212.6_{-1.3}^{+1.4}$ & $232.8_{-1.6}^{+1.7}$ & $231.9_{-1.5}^{+1.4}$ & $235.02_{-0.83}^{+0.86}$ \\ \hline
$R_0$ & \textcolor{magenta}{$8$} & \textcolor{magenta}{$8$} & \textcolor{magenta}{$8$} & \textcolor{magenta}{$8$} & \textcolor{magenta}{$8$} & \textcolor{magenta}{$8$} & \textcolor{magenta}{$8$} \\ \hline
$\alpha_{z}$ & \textcolor{magenta}{$0$} & \textcolor{magenta}{$0.0471$} & \textcolor{magenta}{$0$} & \textcolor{magenta}{$0$} & $0.048_{-0.0018}^{+0.0019}$ & $0.0471_{-0.0019}^{+0.0016}$ & $0.0471_{-0.0019}^{+0.0019}$ \\ \hline
$\alpha_{R}$ & \textcolor{magenta}{$0$} & \textcolor{magenta}{$0$} & \textcolor{magenta}{$0$} & \textcolor{magenta}{$0$} & \textcolor{magenta}{$0$} & \textcolor{magenta}{$0$} & $0.67_{-0.26}^{+0.25}$ \\ \hline
$\chi_{\rm red}^2$ RAVE & 2.07 & 1.80 & 2.40 & 1.52 & 1.43 & 1.42 & 1.42 \\ \hline
$\chi_{\rm red}^2$ GCS & 3.85 & 3.86 & 4.08 & 5.15 & 5.57 & 5.42 & 5.46 \\ \hline
\end{tabular}
\medskip
\end{table*}

\section{Constraints on kinematic parameters}
First, we discuss the fiducial parametric model for the Galaxy developed
a decade ago by \citet{2003A&A...409..523R}.
The so-called Besan\c{c}con model
is based on Gaussian velocity ellipsoid functions.
In the {\sl Galaxia} code, the tabulated functions
of \citet{2003A&A...409..523R} were replaced by analytic expressions,
the parameters of which are given in \tab{tbfid}.
One main difference between the {\sl Galaxia} and Besan\c{c}on models
is the value of $R_{0}$ and the Solar motion with respect to the LSR.
Also, {\sl Galaxia} uses slightly different values of $R_\sigma$.
In the Besan\c{c}on model, the velocity dispersions are assumed
to saturate abruptly at around $\tau_{\rm sat}=6.5$ Gyr.
Moreover,
the velocity dispersion of the thick disc does not have any
radial dependence, hence the value of $R_\sigma^{\rm thick}$ only
contributes to the calculation of the asymmetric drift. Neither
of these Ans\"atze are assumed in our analysis.

Finally, in the Besan\c{c}on model, the metallicity [Fe/H] of the thick
disc is assumed to be $-0.78$ with a spread of 0.3 dex.
The spread is not taken into account when assigning magnitudes
and colors from isochrones.  This was done so as to prevent the
thick disc from having a horizontal branch. We do {\it not} make
this ad hoc assumption. Since our data do not
have a strong color-sensitive selection,
this has a negligible impact on our kinematic study.

We now discuss the results obtained from fitting models to the
RAVE and the GCS data. The best-fit parameters and their uncertainties
obtained using MCMC simulation for different models and data
are shown in \tab{tbgauss} and \tab{tbshu}. Note, the
uncertainties quoted in the table are purely random and do
not include systematics. We discuss systematics separately
in \sec{systematics}.
We begin by discussing results from the Gaussian distribution
function before proceeding to the Shu distribution function.

\subsection{Gaussian models}
First we concentrate on GCS data (column 1 of \tab{tbgauss}).
For GCS we find that all the values are well constrained.
However, percentage wise $R_\sigma^{\rm thin}, R_\sigma^{\rm thick}$ and
$V_{\odot}$ have larger
uncertainties as compared to other parameters.
In \fig{gcs_UVW3}, where fits from column 1 are plotted,
it can be seen that the
model is an acceptable fit to the data.
The reduced $\chi^2$ values are
quite high especially in comparison to the mock models.
This is mainly due to significant amount of structure in $(U,V)$
velocity space (see \fig{gcs_UVW3}).
The $\beta_z$, $\sigma_z^{\rm thin}$, $\sigma_z^{\rm thick}$ and $\sigma_R^{\rm thick}$
parameters are close to  the corresponding Besan\c{c}on values but show
other differences. The most notable differences are that our value
for $R_\sigma^{\rm thin}$ is smaller, $R_\sigma^{\rm thick}$ is longer, and
$\sigma_{\phi}^{\rm thick}$ is lower.
Other minor differences are as follows.
Our $\beta_R$ and $\beta_{\phi}$ are lower
and so are the velocity dispersions $\sigma_R^{\rm thin}$,
$\sigma_{\phi}^{\rm thin}$.
The thin-disc velocity dispersions are strongly
correlated to $\beta$ values, so fixing $\beta$ to higher values
will drive the corresponding thin-disc velocity dispersions closer
to the Besan\c{c}on values.
The second column in \tab{tbgauss} shows the results for the case
where a separate thick disc is not assumed (the
thick-disc stars are labelled as thin-disc
in the model). In this case,
$\beta$, $\sigma$ increase, while  $R_\sigma^{\rm thin}$ decreases,
which is expected since the thin disc  has to accommodate
the warmer thick disc component.

We now discuss results for the RAVE data, beginning with the model
where $\alpha_z=0$ (column 4 of \tab{tbgauss}).
Surprisingly, $R_\sigma^{\rm thin}$ is found
to be negative, whereas the $R_\sigma^{\rm thick}$ is positive.
The value of $\Theta_0$
is found to be significantly less than that reported in literature.
The $\beta_{R}$ and $\beta_{\phi}$ values are also too small.
We note that the $\beta_z$ value in RAVE has more uncertainty
than that in GCS, which we had also noted in the  tests on mock data.
From now on we keep
$\beta_z=0.37$, a value we get from GCS. We checked and found
that fixing $\beta_z$ has negligible impact on other parameters.

We now let $\alpha_z$ free and this results in higher value
of $\Theta_0$. The value of $\Omega_{\odot}$ is now close
to the  proper motion of Sgr A*.
Allowing for a vertical dependence of circular speed
decreases $R_\sigma^{\rm thin}$ while increasing $\beta_{R}$ and  $\beta_{\phi}$.
However, these values are still lower than the GCS values.
It can be seen from red lines in \fig{rave_UVW1} that the model does
not fit well the projected $V$ components of velocity.
Clearly there are some problems with this model.

We now compare RAVE and GCS results using columns 6 and 3, where we fix
$R_\sigma^{\rm thin}$, $R_\sigma^{\rm thick}$ and $\alpha_z$ to values that
we will get later from the Shu model. Having the same value of $R_\sigma$ in
both RAVE and GCS makes it easier to compare the other parameters.
Naturally, fixing some of the variables leads to an increased $\chi_{\rm red}^2$.  We
find that most of the values agree to within 4$\sigma$ of each other. The two
exceptions are $\beta_{\phi}$ and $V_{\odot}$ which are higher for GCS.

To summarize, we find that the model parameters
that best fit the RAVE data show important differences from those from GCS.
The models differ mostly in their
values of $R_\sigma^{\rm thin}$ and $R_\sigma^{\rm thick}$, with the RAVE
values being systematically too high.  If $R_\sigma^{\rm thin}$ and
$R_\sigma^{\rm
thick}$ are fixed to be same, then $V_{\odot}$ in RAVE is found
to be lower by about $2 \kms$.
The values of $\beta_{\phi}$ and $\beta_{R}$ are also slightly
lower in RAVE, and are better constrained than $\beta_z$.

\subsection{Shu models}

\begin{figure*}
\centering \includegraphics[width=0.98\textwidth]{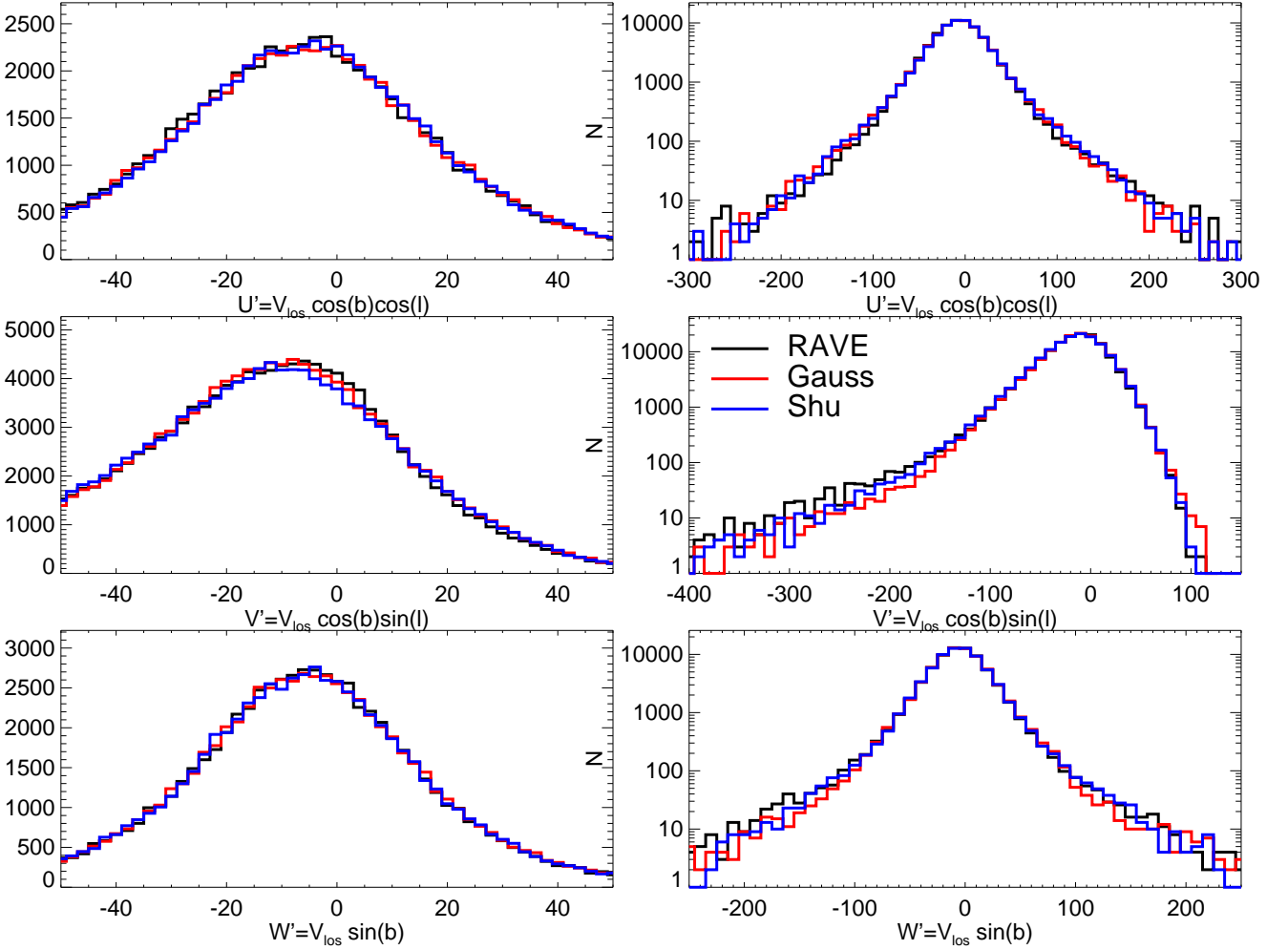}\caption{Comparison of model velocity distributions with that of RAVE data. Projection of radial velocity along $U,V$ and $W$ directions
are shown. The right panels differ from the left only in range
and scale of axes.
The top panel is for stars with $(|b|<45) \& ((|l|>45) || (|l-180| >
45))$,
the middle panel is for stars with
$(|b|<45) \& ((|l|<45) \/|| (|l-180| < 45))$
and the bottom panel is for stars with $|b|>45$.
The model used is the best fit Gaussian (column 5 in
\tab{tbgauss}) and the Shu model (column 6 of \tab{tbshu}) for the
RAVE data. The Shu model clearly models the wings of $V'$ better than
the Gaussian model,  especially in region $-200$ km s$^{-1}$ $<V'<-150$
km s$^{-1}$   and $V'>80$ km s$^{-1}$ which is   dominated by thick disc.
A slight mismatch at $V' \sim 0$ is  also seen.
\label{fig:rave_UVW1}}
\end{figure*}

First, we discuss RAVE results for the case where most of the parameters were
free (column 6 of \tab{tbshu}).  We find that $R_\sigma^{\rm thin}$ is
positive, unlike for the Gaussian model. It can be seen from \fig{rave_UVW1}
that the wings of the $V$ component of velocity are better fitted by the Shu
model than the Gaussian model.  Another important feature is that
$\sigma_{R}$ for the thick disc is almost the same as for the thin disc.
The $\sigma_{z}$ values are also not too far apart.  Apparently, as compared
to Gaussian model, the velocity dispersions for the thick disc are very
similar to that of old thin disc in the Shu model. However, $R_\sigma^{\rm
thick}$ is shorter than $R_\sigma^{\rm thin}$. If $\alpha_z$ is set to zero,
$\Theta_0$ is  underestimated (column 4).  If we impose the measured proper
motion of Sgr A* as a prior, we can constrain the radial gradient of
circular speed, which is found to be less than $1 \kmskpc$ (column 7).
Comparing columns 5 and 6 it can be seen that fixing $\beta_z$ to 0.37
mainly changes $\sigma_z^{\rm thin}$ while the other parameters are
relatively unaffected.

The thick-disc parameters for the GCS sample (column 1 of \tab{tbshu}) differ
significantly from those for the RAVE sample. This is mainly due to the GCS having
very few thick disc stars.  We next fix $R_\sigma^{\rm thick}=7.58\kpc$ and
$R_\sigma^{\rm thin}=13.7\kpc$ for GCS.  Doing so improves the agreement
between the two sets for the thick disc while the change in $\chi_{\rm red}^2$ is very
small (column 2).  Most RAVE parameters agree to within 4$\sigma$ of GCS
except for $V_{\odot}$, which is lower by about $2 \kms$ for RAVE. Finally we
also test models where the thick disc is ignored (column 3). As in the case
of Gaussian models, this leads to an increase in $\beta$, $\sigma$ and
decreasing $R_\sigma^{\rm thin}$.

\begin{figure*}
\centering \includegraphics[width=0.98\textwidth]{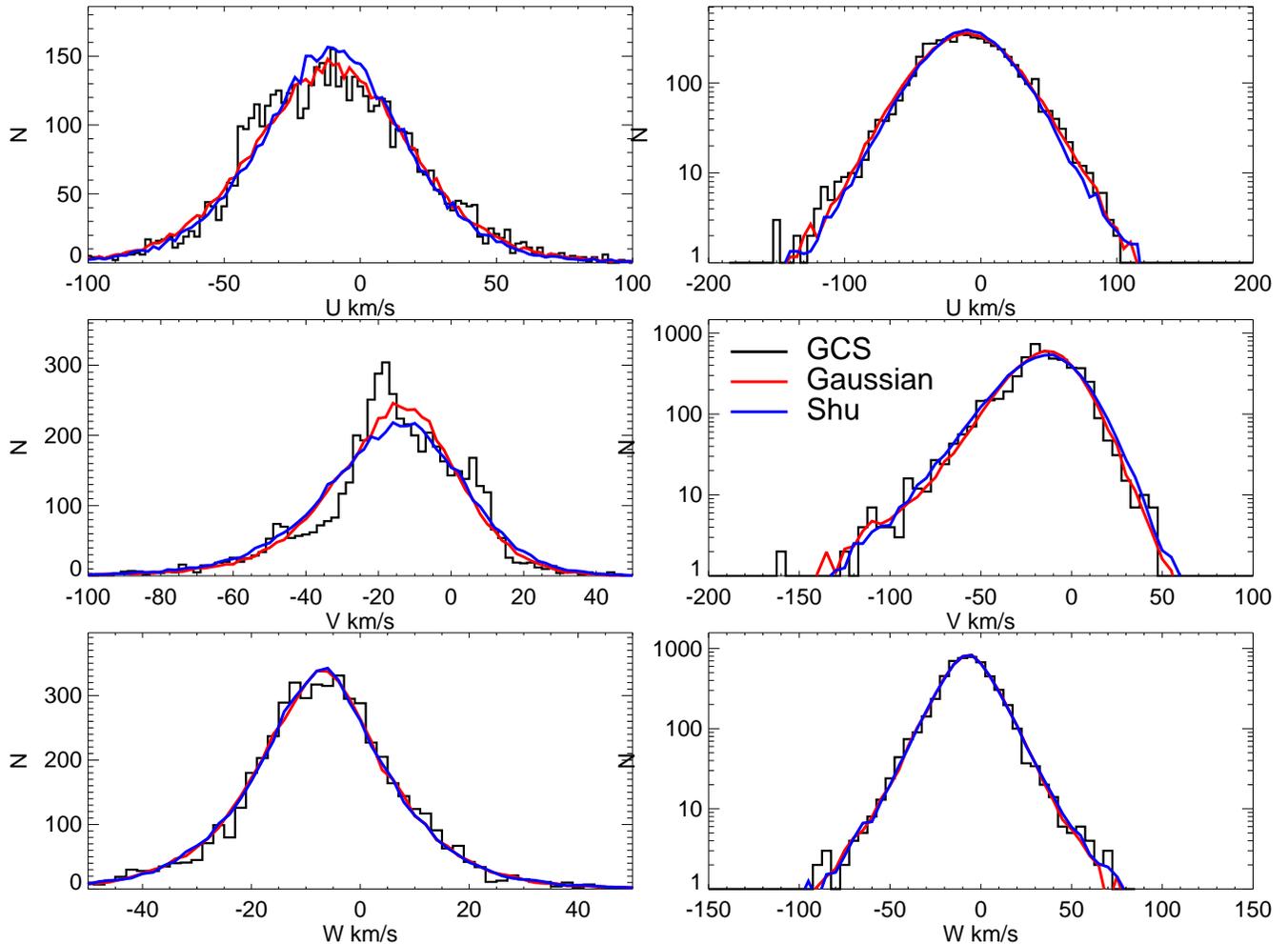}\caption{Comparison of model velocity distributions with that of GCS data. The right panels differ from the left only in range
and scale of axes.
The model used is the best fit Gaussian (column 1 of \tab{tbgauss})
and the Shu model (column 1 of \tab{tbshu}) for the  GCS data.
Both the models are acceptable fits to the data. Significant
structures can be seen in the velocity space.
\label{fig:gcs_UVW3}}
\end{figure*}

\begin{figure*}
\centering \includegraphics[width=0.98\textwidth]{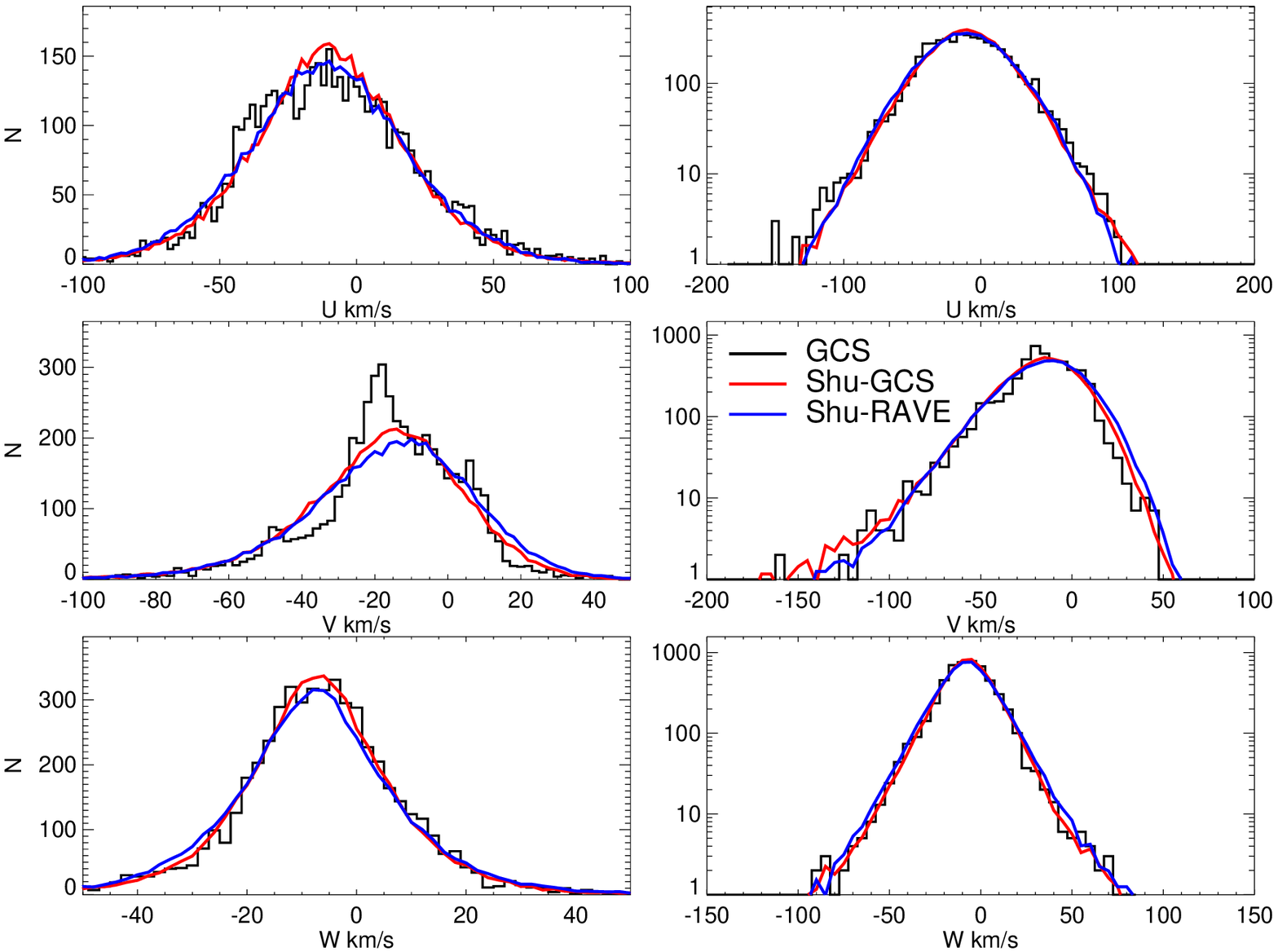}\caption{Comparison of model velocity distributions with that of GCS data. The models used correspond to columns 2 and 6 of \tab{tbshu}.
These are Shu models that a) best fit the
GCS data but with a few parameters fixed and b) best fit the RAVE
data.
The positive wing of $V$ is slightly
overestimated by the RAVE best fit model.
\label{fig:gcs_UVW4}}
\end{figure*}

In \fig{gcs_UVW3} the best fit Gaussian and Shu models
for GCS are compared. Unlike RAVE both models
provide good fits. In fact, to discriminate the models
one requires a large number of warm stars
that can sample the wings of the $V$ distributions
with adequate resolution. The GCS sample
clearly lacks these characteristics.
Next, in \fig{gcs_UVW4} we plot the GCS Shu model alongside the RAVE
Shu model (columns 2 and 6 of \tab{tbshu}) and compare them
with the GCS velocities. It can be seen that both are acceptable
fits. However, the RAVE Shu model slightly
overestimates the right wing of the GCS $V$ distribution.
Note, in \fig{rave_UVW1} a slight mismatch at $V' \sim 0$
can be seen, the cause for this is not yet clear.

\section{Discussion}
\subsection{Correlations and degeneracies}
\begin{figure*}
\centering \includegraphics[width=0.97\textwidth]{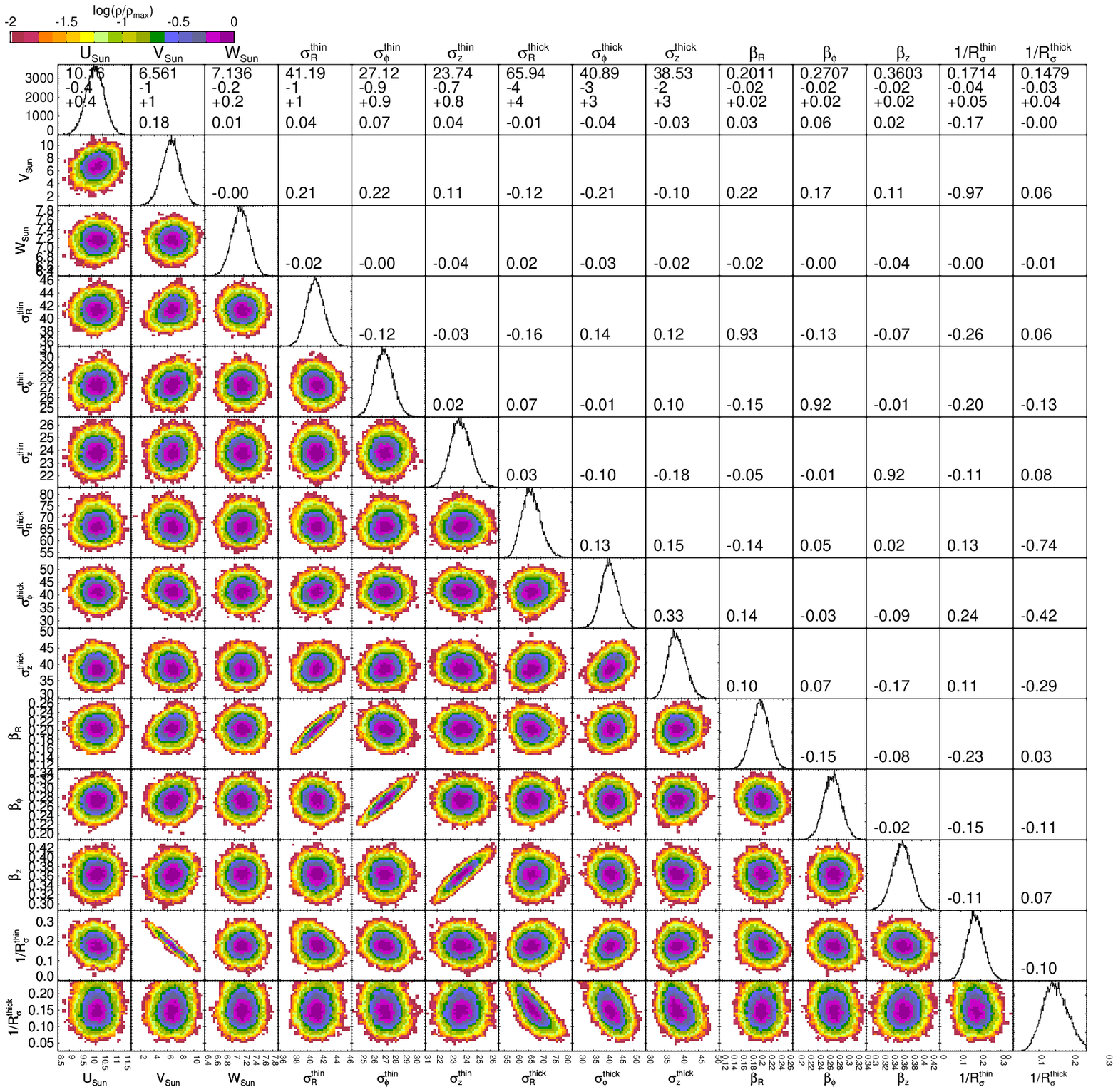}\caption{Marginalized posterior distribution of model parameters.The numbers are  the linear Pearson correlation coefficient.
Shown  is the case of Gaussian model for GCS data (column 1 of \tab{tbgauss}).
Strong dependency can be seen between $\beta$  and $\sigma^{\rm thin}$
values.
Additionally, $(R_{\sigma}^{\rm thin},V_{\odot})$, $(R_{\sigma}^{\rm thin},\sigma_{R}^{\rm thin})$, $(R_{\sigma}^{\rm thin},\beta_{R})$ and $(R_{\sigma}^{\rm thick},\sigma_{R}^{\rm thick})$  also show dependency.
\label{fig:gcs_gauss}}
\end{figure*}

\begin{figure*}
\centering \includegraphics[width=0.97\textwidth]{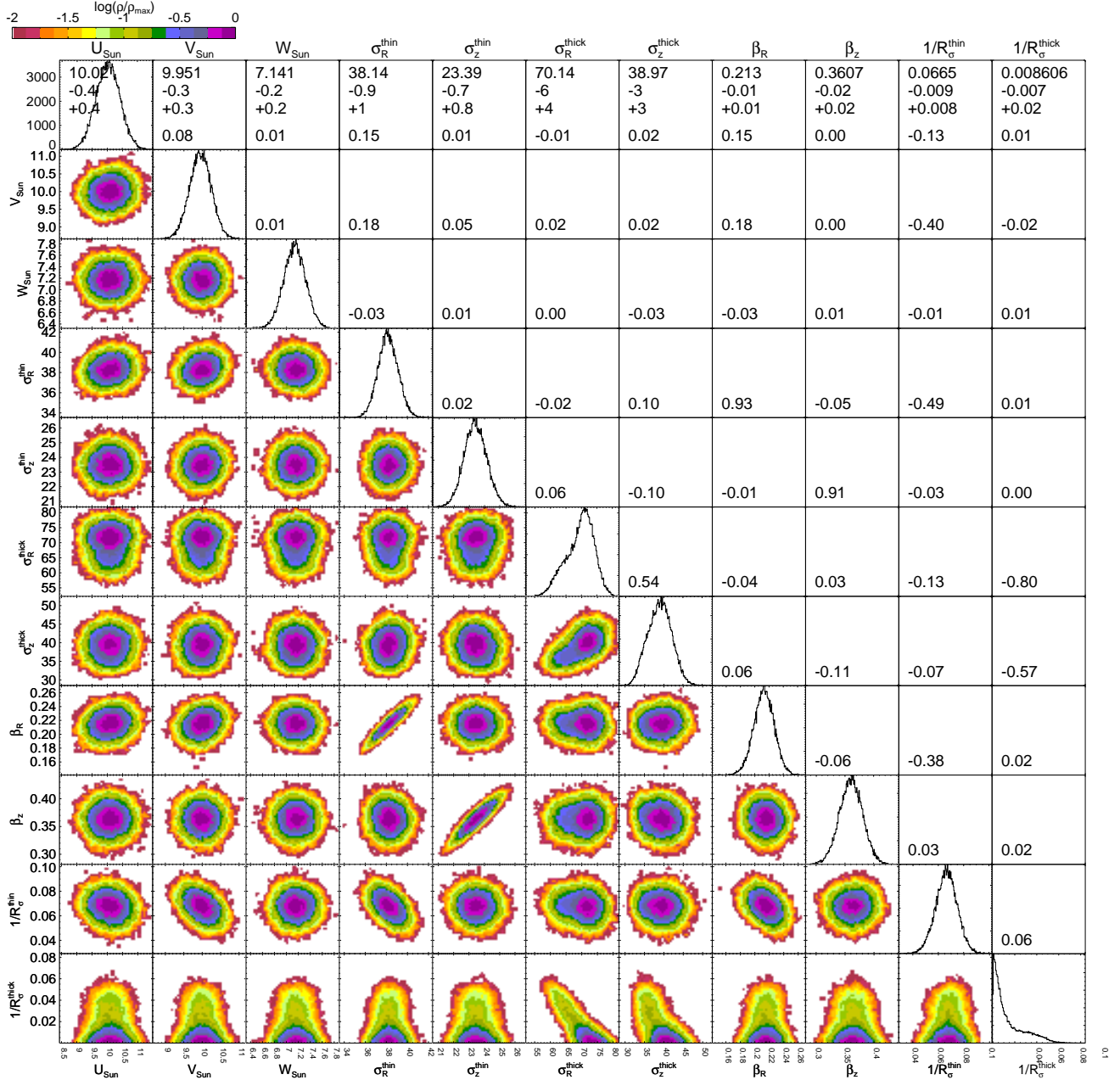}\caption{Marginalized posterior distribution of model parameters.The numbers are  the linear Pearson correlation coefficient.
Shown  is the case of Shu model for GCS data (column 1 of \tab{tbshu}). Same
dependencies as in \fig{gcs_gauss} can be seen.
Dependency of $(R_{\sigma}^{\rm thin},V_{\odot})$ has got weaker while that
of $(R_{\sigma}^{\rm thick},\sigma_{R}^{\rm thick})$ has become stronger.
\label{fig:gcs_shu}}
\end{figure*}

\begin{figure*}
\centering \includegraphics[width=0.97\textwidth]{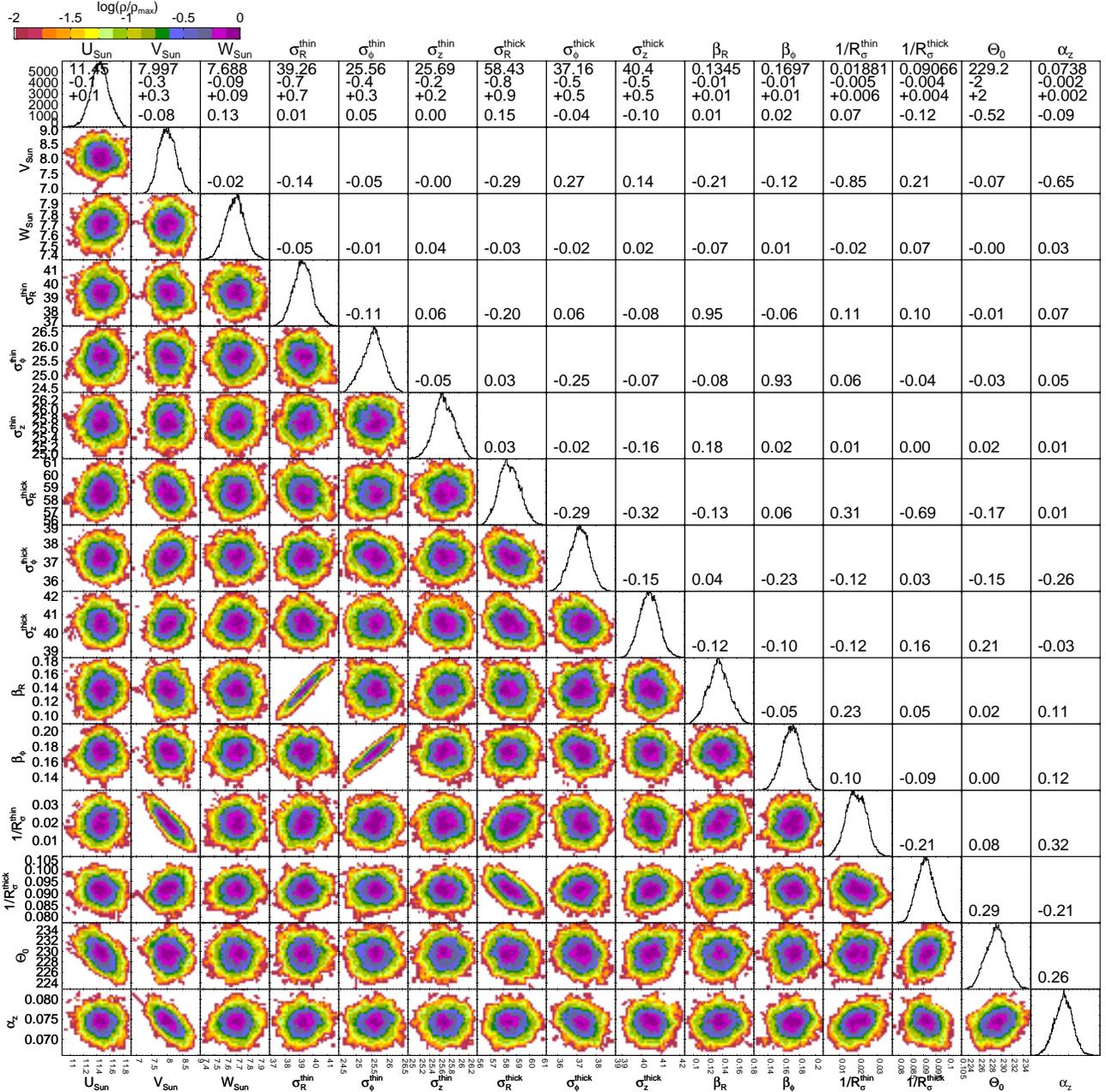}\caption{Marginalized posterior distribution of model parameters.The numbers are  the linear Pearson correlation coefficient.
Shown  is the case of Gaussian model for RAVE data (column 5 of \tab{tbgauss}).
Strong  dependency can be seen between $\beta$  and $\sigma^{\rm thin}$
values.
Additionally, $(R_{\sigma}^{\rm thin},V_{\odot})$ and $(R_{\sigma}^{\rm thick},\sigma_{R}^{\rm thick})$  also show dependency.
Finally, the $\Theta_0$ is anti-correlated to $U_{\odot}$
and $\alpha_z$ to $V_{\odot}$.
\label{fig:rave_gauss}}
\end{figure*}

Not all parameters are independent. The dominant
correlations are shown in Figures \ref{fig:gcs_gauss}, \ref{fig:gcs_shu},
\ref{fig:rave_gauss} and \ref{fig:rave_shu} where pairwise
posterior distributions of parameters are plotted.
The implication of any correlation
is that a change in one of the values also changes the other value
{\it without affecting the quality of the fit.} In other words, a precise
value of one correlated quantity needs to be known in order to determine
the other. We find that the $\beta$ values are strongly correlated with
the corresponding $\sigma^{\rm  thin}$ values. This is mainly
because we do not have enough information in the data
to estimate the ages of stars. The model specifies
the prior on the ages of stars and the data gives
the velocities. The degeneracy reflects the fact
that during fitting $\beta$ can be adjusted while keeping the
mean velocity dispersion constant.

In both thin and thick discs  $\sigma_{R}$ is correlated with $R_\sigma$.  These
correlations are stronger for the Shu model than the Gaussian model.  To get
a good estimate of $R_\sigma$ ideally one would require a sample of stars
distributed over a large volume. In the absence of an extended sample, the
constraint on $R_\sigma$ comes from the fact that it also determines the
$v_\phi$ distribution.  The amount of asymmetric drift increases with
$\sigma_R$ and decreases with $R_\sigma$ (see \equ{stromberg}).  If the
asymmetric drift is fixed, this naturally leads to the correlation between
$R_\sigma$ and $\sigma_R$. In the Shu model the effective velocity dispersion
$\sqrt{\langle v_R^2\rangle}$ is not only proportional to $\sigma_R$ but also
decreases with $R_\sigma$. So one can keep the effective velocity dispersion
constant by decreasing both $R_\sigma$ and $\sigma_R$ at the same time. This
makes the correlation in Shu model stronger.

Also, $V_{\odot}$ is correlated with $R_\sigma^{\rm
thin}$ and this relation is stronger for the Gaussian model.
This makes it difficult to determine $V_{\odot}$ and $R_\sigma$ reliably
using the Gaussian models. The Shu model does not have this problem
because in it the azimuthal motion is coupled to the radial motion,
so it has three fewer parameters, i.e., has fewer degrees of freedom.
This helps to resolve the $R_\sigma^{\rm thin}-V_{\odot}$ degeneracy.

When fitting Shu models to RAVE we find
an anti-correlation
exists between thin and thick disc parameters, e.g.,
$(\sigma^{\rm thin}_{R},\sigma^{\rm thick}_{R})$,
$(\sigma^{\rm thin}_{z},\sigma^{\rm thick}_{z})$ and
$(R_\sigma^{\rm thin},R_\sigma^{\rm thick})$.
This is mainly because we do not have any useful information
about the ages of stars.

We now discuss the parameters $\Theta_0$ and $\alpha_z$ which were free only
for RAVE data.  The value of $\alpha_z$ is correlated with $\Theta_0$ and
anti-correlated with $V_{\odot}$.  The $\Theta_0$ parameter is
anti-correlated with both $U_{\odot}$ and $R_\sigma^{\rm thin}$. For the GCS
data, the $(\Theta_0,R_\sigma^{\rm thin})$ correlation is so strong that it
is difficult to get meaningful constraints on $\Theta_0$ so the later was
fixed.

\subsection{Solar peculiar motion}
Among the three components of Solar motion,  $U_{\odot}$ and
$W_{\odot}$ are only weakly correlated with other variables
and give similar values for both Gaussian and Shu models.
The only major dependence of $U_{\odot}$ is for RAVE, where it is
anti-correlated with $\Theta_0$  by about $-0.5$.
So models with $\alpha_z=0$ that underestimate $\Theta_0$, will
overestimate $U_{\odot}$.
For RAVE we get  $W_{\odot}=7.54\pm 0.1 \kms$  and $U_{\odot}=10.96\pm
0.14\kms$ (column 6 of \tab{tbshu}).
GCS values for $W_{\odot}$ and $U_{\odot}$ are lower
by about 0.4 and 0.8 km s$^{-1}$ respectively but their
$3-\sigma$ range matches with RAVE (column 2 of \tab{tbshu}).
The small mismatch could be either
due to large-scale gradients in the mean motion of stars
\citep{2013MNRAS.436..101W} in RAVE or due to kinematic
substructures in GCS.

Our GCS results (column 2 of \tab{tbshu}) are in excellent agreement with
\citet{1998MNRAS.298..387D}, but differ from \citet{2010MNRAS.403.1829S} for
$U_{\odot}$ by 1.0 km s$^{-1}$.  Nevertheless, $U_{\odot}$ is well within
their quoted $2\sigma$ range. The RAVE $U_{\odot}$ agrees with
\citet{2010MNRAS.403.1829S}.  Interestingly, with the aid of a
model-independent approach, \citet{2012MNRAS.427..274S} finds from SDSS stars
$U_{\odot}=14.0\pm0.3$ km s$^{-1}$ but with a systematic uncertainty of 1.5
km s$^{-1}$.  The systematic errors in distances and proper motion can bias
this result.  Additionally, the analyzed sample not being local, his results
can also be biased if there are large-scale streaming motions.

\begin{figure*}
\centering \includegraphics[width=0.97\textwidth]{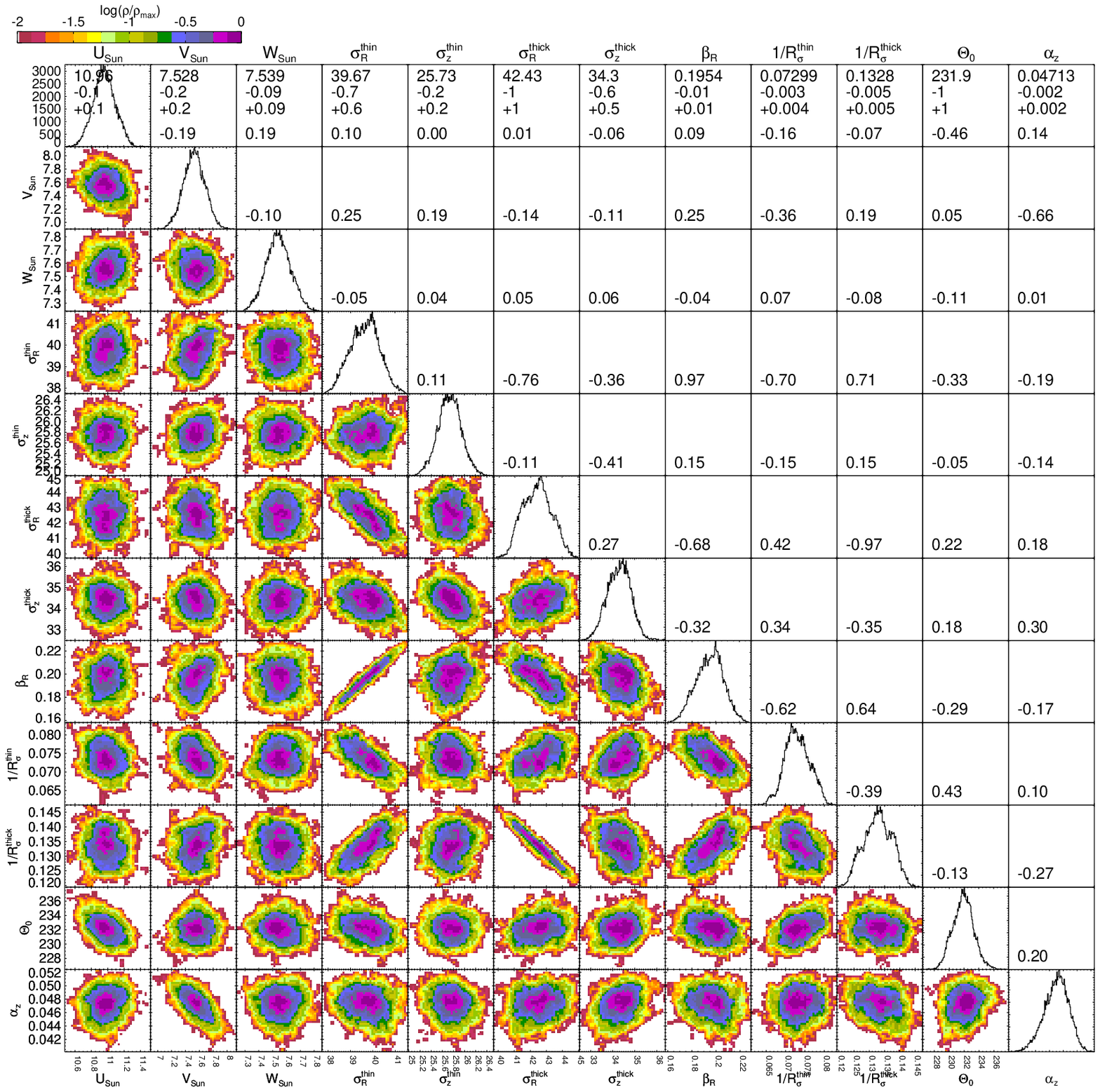}\caption{Marginalized posterior distribution of model parameters.The numbers are  the linear Pearson correlation coefficient.
Shown  is the case of Shu model for RAVE data (column 6 of \tab{tbshu}).
Strong  dependency can be seen between $\beta$  and $\sigma^{\rm thin}$
values.
Additionally, $(R_{\sigma}^{\rm thin},V_{\odot})$, $(R_{\sigma}^{\rm thin},\sigma_{R}^{\rm thin})$, $(R_{\sigma}^{\rm thin},\beta_{R})$ and $(R_{\sigma}^{\rm thick},\sigma_{R}^{\rm thick})$  also show dependency.
Unlike GCS a dependency of $(\sigma_{R}^{\rm
thin},\sigma_{R}^{\rm thick})$ and $(\beta_{z},\beta_{R})$
can be seen. Finally, the $\Theta_0$ is anti-correlated to $U_{\odot}$
and $\alpha_z$ to $V_{\odot}$.
\label{fig:rave_shu}}
\end{figure*}

We now discuss our results for $V_{\odot}$.
For Gaussian models the estimated $V_{\odot}$ value depends
strongly on the choice of $R_\sigma$ values and it is difficult
to get a reliable value for either of them.
For the Shu model, $V_{\odot}$ depends on whether
$\alpha_z$ is  fixed, in fact they are  anti-correlated (see
\fig{rave_shu}).
For $\alpha_z=0$, the GCS and RAVE $V_{\odot}$ agree with
each other, but when $\alpha_z$ is free,
$V_{\odot}$ is $2 \kms$ lower from  RAVE than from GCS (columns 2 and
6 of \tab{tbshu}).
The $\alpha_z=0$ model not only has a higher $\chi_{\rm red}^2$ but,
as we will discuss later, also yields a low value of $\Theta_0$,
so we consider this model less useful.
The most likely cause for the difference between
RAVE and GCS $V_{\odot}$ is the
significant amount of kinematic substructures in the distribution
of the $V$ component of the GCS velocities (\fig{gcs_UVW4}).
It can be seen in \fig{gcs_UVW4} that
the best fit RAVE model, in spite of apparently having low
$V_{\odot}$,  is still a good description
of the GCS data. Moreover, in GCS
a dominant kinematic structure can be seen at $V \sim -20 \kms$
(the Hyades and the Pleiades), lending further support to the idea
that GCS probably overestimates $V_{\odot}$.
However, this
can also be because our formulation for the
vertical dependence of kinematics is not fully self
consistent (see Section \ref{sec:potmodel} and \ref{sec:systematics}).

The need to revise $V_{\odot}$ upwards from the value of $5.2$ km
s$^{-1}$ given by \citet{1998MNRAS.298..387D}
has been extensively discussed
\citep{2010MNRAS.401.2318B,2010MNRAS.402..934M,2010MNRAS.403.1829S}.
\citet{2010MNRAS.401.2318B} suggests a value of $11.0$ km s$^{-1}$ after
randomizing some of the stars to reduce the impact of streams while
\cite{2010MNRAS.403.1829S} get $V_{\odot}=12.24\pm0.47$ km
s$^{-1}$. Our RAVE value of $V_{\odot}=7.5\pm0.2$ is significantly
lower that this (column 6 of \tab{tbshu}).
Our GCS value
of $V_{\odot}=9.8\pm0.3$ km s$^{-1}$ is also lower than both of them
(column 2 of \tab{tbshu}).

Recently, \citet{2013A&A...557A..92G} determined $V_{\odot}=3.06\pm0.68$ by
binning the local RAVE stars in color and metallicity bins and applying an
improved version of the Stromberg relation.  Their estimate is even lower
than that of \citet{1998MNRAS.298..387D}.  The application of the Stromberg
relation demands the identification of subpopulations that are in dynamical
equilibrium and have the same value for the slope in the relation. Binning by
color fails to satisfy these requirements for the reasons given by
\citet{2010MNRAS.403.1829S}.  Golubov et al.\ do split their sample by
metallicity as well as color, but the metallicities are quite uncertain and
the bins are quite broad, so a bias due to the selected subpopulations not
obeying the same linear relation can be expected.

The discrepancy for the GCS with
\citet{2010MNRAS.403.1829S} could be either due to
differences in fitting methodologies or differences in the models
adopted, with the latter being the most likely cause.
The model used here and by \citet{2010MNRAS.403.1829S} is based on the
Shu distribution
function but still there are some important differences.
We have a separate thick disc
while in their case the thick disc arises naturally due to
radial mixing. The forms of $\sigma_R(L)$ and $\Sigma(L)$
also differ ($L$ being angular momentum).
Our form of $\sigma_R(L)$ is the same as that
used by \cite{2010MNRAS.401.2318B} while \cite{2010MNRAS.403.1829S}
compute $\sigma_R(L)$ so as to satisfy
$\langle v_{R,{\rm thin}}^2\rangle \propto {\rm e}^{-R/1.5R_d}$. In our case $\langle v_{R,{\rm thin}}^2\rangle(R)$
depends implicitly upon $R_\sigma$ and $\beta$ and both of these parameters are
constrained by data. The $\Sigma(L)$ in \cite{2010MNRAS.403.1829S}
comes from a numerical simulation involving the processes of
accretion, churning and blurring while in our case it comes directly
from the constraint that $\Sigma(R) \propto \exp(-R/R_d)$.
The prescription for metallicity in  \cite{2010MNRAS.403.1829S}
is also very different from ours.

\subsection{The circular speed}
In a recent paper, \citet{2012ApJ...759..131B} used data from the APOGEE
survey and analyzed stars close to the mid-plane of the disc to find
$\Theta_0=218\pm 6$ km s$^{-1}$ and $V_{\odot}=26\pm3\kms$.
The resulting angular velocity
$\Omega_{\odot}=(\Theta_0+V_{\odot})/R_{0}$ agrees with the value
of $30.24\pm0.11\kms\kpc^{-1}$ as estimated by
\citet{2004ApJ...616..872R} using the Sgr A*
proper motion or as estimated by \citet{2010MNRAS.402..934M} using
masers ($\Omega_{\odot}$ in range $29.9-31.6\kms\kpc^{-1}$).
However, \citet{2012ApJ...759..131B} found that $V_{\odot}$ is about $14\kms$
larger than the value measured in the solar neighborhood by GCS.
As a way to reconcile their high $V_{\odot}$,
\citet{2012ApJ...759..131B} suggest that the LSR itself is rotating
with a velocity of $\sim 12\kms$ with respect to the RSR (rotational
standard of rest as measured by circular speed in an axis-symmetric
approximation of the full potential of the Milky Way).

For RAVE data, we get $\Theta_0=232 \pm 1.7\kms$
and $\Omega_{\odot}=29.9\pm 0.3 \kmskpc$
which agrees with the proper motion of Sgr A*,  $30.24 \pm 0.1\kmskpc$.
Hence, the RAVE data suggest that the LSR is on a circular
orbit and is consistent with RSR.
Our value $\alpha_z=0.047$ is slightly higher than the
value $0.0374$ predicted by analytical models
of the Milky Way potential \fig{vcirc_z}. This is expected
because in our formalism,
the parameter $\alpha_z$ also contributes to the decrease in mean rotation
speed  with height.
If we explicitly put a prior on
$\Omega_{\odot}$, then we have the liberty of constraining one more
parameter and we use it to constrain the radial gradient of circular
speed $\alpha_R$. Doing so, we find a small gradient of about
$0.67 \kmskpc$  (column 7 of \tab{tbshu}) and $\Theta_0$ increases to $235
\kms$.

We find that the parameter $\alpha_z$ that controls the vertical
dependence of circular speed plays an important role in
determining $\Theta_0$.
For models with $\alpha_z=0$, $\Theta_0$ is underestimated and we
end up with $\Theta_0=212 \pm 1.4$.
This is in rough agreement with \citet{2012ApJ...759..131B}
but $V_{\odot}$ is not.
The resulting angular velocity
$\Omega_{\odot}$ is also much lower
than the value obtained from the proper motion of Sgr A*.
If, on the contrary, $\alpha_z$ is free, we automatically
match the proper motion of Sgr A* and we get a value of $V_{\odot}$ that
is similar to that from the local GCS sample.

\begin{figure}
\centering \includegraphics[width=0.48\textwidth]{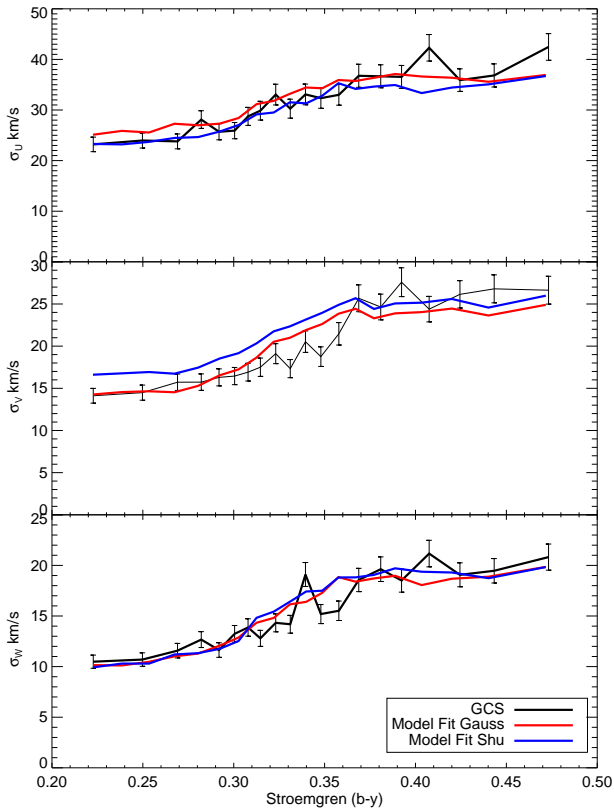}\caption{ Velocity as a function of $b-y$ Str\"omgren color for GCS stars. The  error bars were estimated from Poisson noise.
Shown alongside are predictions from various models.  Note, the color
distribution was not taken into account when fitting models to data.
\label{fig:gcs_colvel}}
\end{figure}

\subsection{The age-velocity dispersion relation (AVR)}

We now discuss our model predictions for the age velocity dispersion relation
in the thin disc, specifically the parameters $\beta_{z},\beta_{\phi},
\beta_{R}$, $\sigma_{z}^{\rm thin},\sigma_{\phi}^{\rm thin}$ and
$\sigma_{R}^{\rm thin}$.  We find $\beta_R < \beta_{\phi} < \beta_z$.  The
GCS $\beta_{R,\phi,z}$ values were similar for both Gaussian and Shu models.
The RAVE value of $\beta_R$ from the Shu model,  also agrees with these GCS
values.  The value of $\beta_z$ is difficult to determine precisely with RAVE,
so, we used the corresponding GCS value in the
fits.  The values of $\beta_{R,\phi}$ from RAVE with the Gaussian model are
systematically lower than the GCS values.  Since the RAVE Gaussian model did
not fit the data well, we give less importance to its $\beta$ values and
ignore them for the present discussion.  Overall, results in column 1 of
\tab{tbgauss} provide a good representation of our predictions and are shown
alongside literature values in \tab{tblit}.

Our values of $\beta$ and the velocity dispersion in the Solar neighborhood
for 10 Gyr old stars, $\sigma_{R,\phi,z}^{\rm thin}$, depend on whether the
thick disc is considered  a distinct component: when only one component is
provided, so the thick disc has to be accommodated by the old tail of the
thin disc, these quantities are naturally higher (column 2 of \tab{tbgauss}).
The values we recover for $\sigma_{R,\phi,z}^{\rm thin}$ are very similar
regardless of which survey or which model we employ.

We now compare our results with previous estimates.  In the Besan\c{c}on
model, the age-velocity dispersion relation for the thin disc was based on an
analysis of Hipparcos stars by \citet{1997ESASP.402..621G}.
\citet{2011ApJ...730....3S} fitted their tabulated values using analytical
functions and the values are given in \tab{tbfid}.
\citet{2004A&A...418..989N} used their ages for individual GCS stars to find
$(\beta_R,\beta_{\phi},\beta_z)=(0.31,0.34,0.47)$.
\citet{2007MNRAS.380.1348S}, using the same data, concluded that the error
bars need enlarging and pointed out that excluding the Hercules stream
increases $\beta_z$ to 0.5.  \citet{2007A&A...475..519H} and
\citet{2009A&A...501..941H} updated the data with new parallaxes and
photometric calibrations and found
$(\beta_R,\beta_{\phi},\beta_z)=(0.39,0.40,0.53)$.  By contrast,
\citet{2010MNRAS.402..461J} used a selection of Hipparcos stars and an
elaborate model of the solar cylinder to estimate $\beta_z=0.375$.
\citet{2009MNRAS.397.1286A} analyzed revised Hipparcos data with a refinement
of the approach of \citet{2000MNRAS.318..658B}. Their analysis used only the
variation with color of velocity dispersion and number density; they did not
use age estimates for individual stars. The advantage of this approach is
that one can include main-sequence stars with colors that span a much wider
range than the GCS catalogue does. The disadvantage is that only proper
motions can be used.  They found $(\beta_R,\beta_{\phi},\beta_z)
=(0.307,0.430,0.445)$.  Since they did not distinguish the thick disc, their
$\beta$ values are closer to the values ($0.268,0.349,0.432$) we obtain  without a
thick disc. For the velocity dispersions, however,
\citet{2009MNRAS.397.1286A} find $(\sigma_R^{\rm thin},\sigma_{\phi}^{\rm
thin},\sigma_z^{\rm thin})=(41.90,28.82,23.83)$, which agree better with our
values when we include a thick disc.

As (\tab{tblit}) shows, our values for $\beta$ are slightly lower than those
from previous studies when we do not include a thick disc, and significantly
lower when a thick disc is included.  While uncertainty in ages remains a big
worry in the analysis of \citet{2009A&A...501..941H}, the difference between
our results with those of \citet{2009MNRAS.397.1286A} is most likely due to
different methods.  The main differences being that we use many fewer stars
stars and we use line-of-sight
velocities rather than proper motions. Also the density laws assumed for the distribution of stars in
space are different.  In \fig{gcs_colvel} we show the velocity dispersion as
a function of Str\"omgren $b-y$ color. Although we have not used this color,
our fitted model correctly reproduces dispersion as a function of color.  The
Shu model is found to overpredict $\sigma_V$ for $(b-y)<0.35$ but only
slightly.

\begin{table}
\caption{\label{tab:tblit}
Comparison of values of $\beta$ as estimated by different sources}
\centering
\begin{tabular}{|l|l|l|l|} \hline
Source & $\beta_R$ & $\beta_{\phi}$ & $\beta_z$ \\ \hline
Fit to Robin et al. (2003) & 0.33 & 0.33 & 0.33 \\ \hline
Nordstrom et al. (2004) & $0.31\pm 0.05$ & $0.34\pm 0.05$ & $0.47 \pm 0.05$ \\ \hline
Seabroke \& Gilmore (2007)  & & & $0.48 \pm 0.26$ \\ \hline
Holmberg et al. (2007) & 0.38 & 0.38 & 0.54 \\ \hline
Holmberg et al. (2009) & 0.39 & 0.40 & 0.53 \\ \hline
Aumer \& Binney (2009) & 0.307 & 0.430 & 0.445 \\ \hline
Just and Jahreiss (2010)  & & & 0.375 \\ \hline
Our GCS Thin only  & $0.27\pm$0.02 & 0.35$\pm$0.02 & 0.43$\pm$0.02 \\ \hline
Our GCS Thin+Thick  & 0.20$\pm$0.02 & 0.27$\pm$0.02 & 0.36$\pm$0.02 \\ \hline
Our RAVE Thin+Thick  & 0.19$\pm$0.01 &  & 0.3-0.4 \\ \hline
\end{tabular}
\end{table}

\begin{figure}
\centering \includegraphics[width=0.38\textwidth]{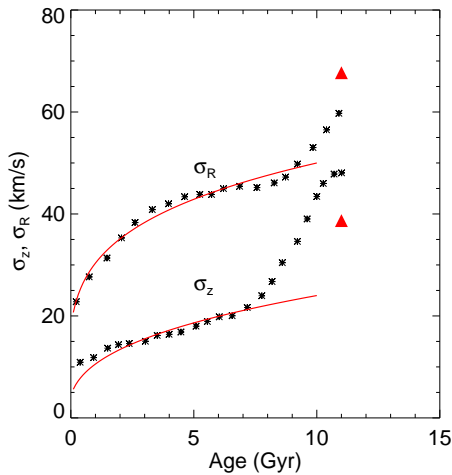}\caption{Comparison of our age-velocity dispersion relation (solid line) with that of \citet{2013A&A...558A...9M} (black points). The
slopes used are $\beta_z=0.37$ and $\beta_R=0.23$ for $\sigma_R=50.0$
and $\sigma_z=24.0$. The triangles are for the thick disc in our
Gaussian models for GCS.
\label{fig:minchev_sigma}}
\end{figure}

The ratio of $\sigma_z/\sigma_R$ and the $\beta_i$ values are useful for
understanding the physical processes responsible for heating the disc.
\citet{1953ApJ...118..106S} first showed that scattering of stars by gas
clouds can cause velocity dispersion to increase with age. This process was
extensively analysed by \citet{1988MNRAS.230..597B}, but they predicted a value of
$\sigma_z/\sigma_R$ from cloud scattering that is too large because they
assumed that an isotropic distribution of star-cloud impact parameters.  When
the anisotropy of impact parameters is taken into account, in the steady
state $\sigma_R/\sigma_z=0.62$.
\citep{1993MNRAS.263..875I,1999MNRAS.307..737S,2008ASPC..396..241S}.
\citet{2002MNRAS.337..731H} showed that with giant molecular clouds one gets
$\beta_R=0.2$ and $\beta_z=0.25$, compared to our favoured values
$\beta_R=0.20$, $\beta_z=0.36$. However, the population of massive gas clouds
is not numerous enough to account for the measured acceleration of thin-disc
stars -- the role of clouds must be to convert random motion in the plane
into random motion vertically
\citep{1992MNRAS.257..620J,2002MNRAS.337..731H}.

\citet{1985ApJ...299..633L} and \citet{2002MNRAS.337..731H} have investigated
scattering by $\sim10^7 M_{\odot}$ halo objects such as black holes and find
then that $\beta_{R,z} \sim 0.5$ and that $\sigma_z/\sigma_R$ lies between
0.40 and 0.67.  Massive halo objects act differently from GMCs for several
reasons: they are not confined to the disc, they are on highly non-circular
orbits, and they have large escape velocities, so they can scatter through
large angles.

For RAVE, from either the Gaussian or Shu models, we get
$\sigma_z^{\rm thin}/\sigma_R^{\rm thin} ~ 0.65$
(column 6 of \tab{tbgauss} and column 6 of \tab{tbshu}).
The corresponding GCS value is 0.58 (column 3 of \tab{tbgauss} and
column 2 of \tab{tbshu}). Models without a thick disc
give a similar value for $\sigma_z/\sigma_R$.
These are values for a 10 Gyr old
population and we think they agree well with the above predictions.
For the thick disc we find that the Gaussian model predicts
$\sigma_z^{\rm thick}/\sigma_R^{\rm thick}=0.68$, while the
Shu model predicts a higher value, 0.80.

Heating by cloud scattering predicts $\beta_R\sim\beta_z$.
Scattering by spiral arms at  Lindblad resonances also heats discs. If spiral
arms are transient, individual  resonances are broad, and over the life of
the disc one or more resonances is likely to have affected every region of
the disc. Spirals
only increase in-plane dispersions
\citep{1985ApJ...292...79C,1988MNRAS.230..597B,2013ApJ...769L..24S}. The predicted values
of $\beta_R$ are between 0.2 for high-velocity stars and 0.5
for low-velocity stars.
Multiple spiral density waves \citep{2006MNRAS.368..623M} or a
combination of bar and spirals
can also heat up the disc \citep{2010ApJ...722..112M}.
When the $\beta_i$ differ from one another, as we find,
the axial ratios of the velocity ellipsoid are functions of age.
If $\beta_z>\beta_R$, $\sigma_z/\sigma_R$ increases
with age as $\tau^{\beta_z-\beta_R}$, so it is much
lower for younger stars. \citet{2009MNRAS.397.1286A} also find that
$\sigma_z/\sigma_R$ increases with age and remark that this trend is
consistent with scattering by spiral arms playing a significant role for
young stars.

Recently, \citet{2013A&A...558A...9M} investigated the
age-velocity dispersion relation for stars in simulations
of disc galaxies and find it to be in rough agreement
with observations. We now compare our results with their
findings. In \fig{minchev_sigma}, we plot their predictions
for $\sigma_R$ and $\sigma_z$  for stars in
a Solar cylinder defined by $7<R<9$ kpc. The red curves show
our AVR from \equ{veldisp1} with $\beta_z=0.37$ and $\beta_R=0.23$, values that
fit both the RAVE and GCS data well when using the Shu model
(column 2 of \tab{tbshu}).
It can be seen that for ages less
than 7 Gyr, the adopted
$\beta$ values correctly reproduce the  profiles
seen in simulations. However, the simulations require a smaller value
$\sigma_z/\sigma_R\sim0.5$ than the data require, and the red curves in
\fig{minchev_sigma} have been individually scaled to fit the simulations.
Hence, although the normalization constant $\sigma_z^{\rm thin}$
is roughly in agreement with our results for the Galaxy, the
normalization constant $\sigma_R^{\rm thin}$
is too high by about $10\kms$.
There is a slight hint that in the simulations
$\sigma_R$ flattens beyond 5 Gyr, but it is also
consistent with our power law prescription.
Since the simulation data are for $7<R<9$ kpc, and the density
of stars and the velocity dispersion increases inwards,
the dispersions in the simulations are expected to be slightly
high compared to dispersions at $R=R_{0}$.
In our model the thin disc started forming 10 Gyr ago (solid line)
and stars older than this belong to the
thick disc with a constant age of 11 Gyr
(shown by red triangles). This is an effective if rather crude
representation of what is found in the simulations.

\subsection{The thick disc}
First, we discuss our results for the Gaussian model.  Our values for
$(\sigma_R^{\rm thick},\sigma_{\phi}^{\rm thick},\sigma_z^{\rm thick})$ for
the thick disc from fitting the Gaussian model to GCS (column 3 of
\tab{tbgauss}) are in good agreement with results of
\citet{2003A&A...398..141S} ($39\pm4,39\pm4,63\pm6$) but differ from those of
\citet{2003A&A...409..523R} regarding $\sigma_{\phi}^{\rm thick}$.  The RAVE
$\sigma_R^{\rm thick}$ is lower than GCS by $7 \kms$ (column 6 of
\tab{tbgauss}) but the other dispersions match up with GCS.

In the Gaussian model the thick disc velocity dispersions
are much larger than those of the old thin disc.
In the Shu models,  we find that
the thick disc dispersions are very similar to the old thin disc
(column 6 of \tab{tbshu}).
However, $R_\sigma^{\rm thick}$ is much shorter than $R_\sigma^{\rm thin}$.
The  Gaussian and Shu models differ in their
estimates for the thick disc velocity dispersions for the following reason.
In the Shu model, the parameter $\sigma_R^2$,
which controls the velocity dispersion, is a function of age
$\tau$ and guiding radius $R_g$ and is
not equal to the velocity dispersion $\overline{v_R^2}(\tau,R)$.
For a positive $R_\sigma$,
$\overline{v_R^2}(\tau,R)=\int \sigma_R^2(\tau,R_g)P(R_g|R,\tau)dR_g
>\sigma_R^2(\tau,R_g=R)$. In a warm
disc there are generally a significant number of stars with $R_g<R$ at radius
$R$. Decreasing
$R_\sigma$ not only makes stars at small radii hotter, but also makes
them more likely to be found at $R>R_g$, so decreasing
$R_\sigma$ increases $\overline{v_R^2}(\tau,R)$.
For the set of parameters given in
column 6 of \tab{tbshu}, we find that at $R=R_0$
\be
\sqrt{\langle v_{z,{\rm thin}}^2\rangle(\tau)} & = & 26.8
\left(\frac{\tau+0.1}{10.1 \Gyr}\right)^{0.41} \kms  \\
\sqrt{\langle v_{R,{\rm thin}}^2\rangle(\tau)} & = & 41.4
\left(\frac{\tau+0.1}{10.1 \Gyr}\right)^{0.22} \kms  \\
\sqrt{\langle v_{z,{\rm thick}}^2\rangle} & = & 40.0 \kms, \\
\sqrt{\langle v_{R,{\rm thick}}^2\rangle} & = & 49.4 \kms,
\label{equ:shu_Solar_disp}
\ee
with $0<\tau<10 \Gyr$.
So the total thick disc $\overline{v_{z,R}^2}$ in the solar neighborhood
is still much larger than that of the thin-disc.

In the Shu model the dispersions at $R_g=R_0$ of the old thin disc and the
thick disc are similar, consistent with the thick disc being merely the tail
of the thin disc. Moreover, although $R_\sigma^{\rm thick}$ is much smaller
than $R_\sigma^{\rm thin}$, we cannot at this stage exclude a smooth decrease
in $R_\sigma$ with age.  Additionally, our prior on age and distance
distribution assumes a distinct thick disc, e.g., in \fig{rave_stat1} it can
be seen that the distance distribution changes suddenly at $10\Gyr$.  This
could be responsible for $R_\sigma^{\rm thick}$ being shorter than
$R_\sigma^{\rm thin}$, perhaps because all scale lengths decrease with age as
\citet{2012ApJ...753..148B} infer.

\begin{table*}
\caption{\label{tab:tbbovy}
Constraints on model parameters with \citet{2012ApJ...759..131B}
Gaussian model. See
\tab{tbgauss} for further description.}
\centering
\begin{tabular}{|l|l|l|l|l|} \hline
Model & RAVE BOVY & RAVE BOVY & RAVE BOVY & RAVE BOV \\ \hline
$U_{\odot}$ & $10.16_{-0.15}^{+0.15}$ & $11.78_{-0.15}^{+0.15}$ & $11.59_{-0.14}^{+0.15}$ & $10.96_{-0.14}^{+0.14}$ \\ \hline
$V_{\odot}$ & $13.36_{-0.22}^{+0.25}$ & $6.2_{-0.18}^{+0.18}$ & $8.77_{-0.28}^{+0.28}$ & $0.032_{-0.024}^{+0.052}$ \\ \hline
$W_{\odot}$ & $7.364_{-0.098}^{+0.098}$ & $7.688_{-0.09}^{+0.09}$ & $7.694_{-0.089}^{+0.097}$ & $7.622_{-0.087}^{+0.094}$ \\ \hline
$\sigma_{\phi}^{\rm thin}$ & $26.455_{-0.096}^{+0.095}$ & $33.11_{-0.27}^{+0.29}$ & $25.83_{-0.38}^{+0.36}$ & $33.84_{-0.27}^{+0.28}$ \\ \hline
$\sigma_R^{\rm thin}$ & $41.39_{-0.16}^{+0.17}$ & $57.58_{-0.31}^{+0.31}$ & $41.55_{-0.62}^{+0.57}$ & $51.44_{-0.27}^{+0.29}$ \\ \hline
$\sigma_z^{\rm thin}$ & $22.99_{-0.12}^{+0.12}$ & $31.55_{-0.3}^{+0.3}$ & $23.29_{-0.62}^{+0.6}$ & $33.6_{-0.29}^{+0.29}$ \\ \hline
$\sigma_{\phi}^{\rm thick}$ &  &  & $37.45_{-0.56}^{+0.5}$ &  \\ \hline
$\sigma_R^{\rm thick}$ &  &  & $65.69_{-0.64}^{+0.55}$ &  \\ \hline
$\sigma_z^{\rm thick}$ &  &  & $38.84_{-0.52}^{+0.47}$ &  \\ \hline
$\beta_R$ & \textcolor{magenta}{$0.01$} & $0.4584_{-0.0066}^{+0.0071}$ & $0.193_{-0.012}^{+0.013}$ & $0.3568_{-0.0049}^{+0.0045}$ \\ \hline
$\beta_{\phi}$ & \textcolor{magenta}{$0.01$} & $0.3747_{-0.009}^{+0.0093}$ & $0.166_{-0.013}^{+0.013}$ & $0.4151_{-0.0093}^{+0.0081}$ \\ \hline
$\beta_z$ & \textcolor{magenta}{$0.01$} & $0.514_{-0.013}^{+0.016}$ & $0.263_{-0.027}^{+0.025}$ & $0.588_{-0.016}^{+0.013}$ \\ \hline
$1/R_{\sigma}^{\rm thin}$ & $-0.029_{-0.0035}^{+0.003}$ & $0.0493_{-0.0024}^{+0.0026}$ & $0.0116_{-0.005}^{+0.0051}$ & $0.0418_{-0.0031}^{+0.0028}$ \\ \hline
$1/R_{\sigma}^{\rm thick}$ &  &  & $0.069_{-0.0039}^{+0.0038}$ &  \\ \hline
$\Theta_0$ & $210.8_{-1.5}^{+1.5}$ & $205.5_{-1.5}^{+1.5}$ & $213.9_{-1.6}^{+1.6}$ & $239.1_{-1.9}^{+1.7}$ \\ \hline
$k_{\rm ad}$ & \textcolor{magenta}{$0.85$} & \textcolor{magenta}{$0.85$} & \textcolor{magenta}{$0.85$} & $1.968_{-0.021}^{+0.022}$ \\ \hline
\end{tabular}
\end{table*}

\subsection{The radial gradient of velocity dispersions}

To date, there has been little discussion in the literature about the
parameter $R_\sigma$ that controls the radial dependence of
velocity dispersion.  This choice of the
radial dependence is motivated by the desire to produce discs in which the
scale height is independent of radius.  For example, under the epicyclic
approximation, if $\sigma_z/\sigma_R$ is assumed to be constant, then the
scale height is independent of radius for $R_\sigma=2R_d$
\citep{1982A&A...110...61V,1988A&A...192..117V,2011ARA&A..49..301V}.
\citet{1989AJ.....97..139L} using 600 old disc K giants spanning 1 to 17 kpc
in galactocentric radius estimate $R_\sigma$ to be $8.7\kpc$ for radial velocity
and $6.7\kpc$
for azimuthal velocity. \citet{1996A&A...311..456O} using a survey of UBVR
photometry and proper motions in different directions of the Galaxy estimated
$R_\sigma=11\pm1.6\kpc$.  \citet{2012ApJ...755..115B} using SDSS/SEGUE data find
$R_\sigma=7.1\kpc$ for vertical velocity dispersions.  \citet{2012ApJ...759..131B}
using APOGEE data find $R_0/R_\sigma$ to be between -0.24 to 0.03, for the radial and
azimuthal motion.  In our modelling, the radial gradient is assumed to be
same for all the three components.

Our results indicate that for GCS, $R_\sigma$ is positive for both Gaussian
and Shu models.  In the case of RAVE, the Shu model yields $R_\sigma
\sim 14\kpc$ but the Gaussian model requires $R_\sigma$ to be negative.
Moreover, we find that when the Gaussian model used by
\citet{2012ApJ...759..131B} is fitted to the RAVE data, $R_\sigma$ is again
negative: $R_\sigma^{\rm thin}=-34\kpc$ (column 1 of \tab{tbbovy}), similar
to their result $(-0.24 <R_0/R_\sigma<0.03)$.  Since the Shu model also fits
the data better, we think that negative values of $R_\sigma$ obtained with
Gaussian models are spurious. {\it The Gaussian model does not fit the RAVE
data well because in a warm disc the $v_{\phi}$ distribution is very skew,
and the Shu DF correctly handles the asymmetry.} Moreover, the $R_\sigma^{\rm
thin}$ estimate from the Shu model agrees for both GCS and RAVE, lending
further support to the proposition that the problem is related to the use of
the Gaussian model.

Positive $R_\sigma$ agrees with the findings of \citet{1989AJ.....97..139L}.
It should be noted that in both our analysis and that of
\citet{2012ApJ...759..131B}, the value of $R_\sigma$ is strongly influenced
by how the asymmetric drift is modelled.  On the other hand, the values
reported by \citet{1989AJ.....97..139L} are a direct measure of the radial
gradient of velocity dispersion.  From RAVE data, the thick disc's value of
$R_\sigma$ is in general higher than the thin disc's value.

\begin{table*}
\caption{\label{tab:shu_alpha}
Investigation of systematics.}
\centering
\begin{tabular}{|l|l|l|l|l|l|l|l|} \hline
Model & RAVE SHU & RAVE SHU & RAVE SHU & RAVE SHU & RAVE SHU & RAVE SHU & RAVE SHU \\ \hline
Distance Change &  & &  &  & 90\% & 110\% & \\ \hline
$U_{\odot}$ & $10.96_{-0.13}^{+0.14}$ & $11.05_{-0.16}^{+0.15}$ & $10.81_{-0.14}^{+0.15}$ & $10.98_{-0.15}^{+0.14}$ & $11.01_{-0.14}^{+0.13}$ & $10.82_{-0.14}^{+0.15}$ & $10.71_{-0.14}^{+0.14}$ \\ \hline
$V_{\odot}$ & $7.53_{-0.16}^{+0.16}$ & $7.62_{-0.16}^{+0.13}$ & $7.39_{-0.14}^{+0.14}$ & $7.59_{-0.14}^{+0.16}$ & $8.26_{-0.15}^{+0.15}$ & $6.81_{-0.16}^{+0.15}$ & $7_{-0.16}^{+0.15}$ \\ \hline
$W_{\odot}$ & $7.539_{-0.09}^{+0.095}$ & $7.553_{-0.09}^{+0.086}$ & $7.52_{-0.088}^{+0.085}$ & $7.535_{-0.089}^{+0.082}$ & $7.553_{-0.091}^{+0.078}$ & $7.53_{-0.083}^{+0.09}$ & $7.517_{-0.088}^{+0.088}$ \\ \hline
$\sigma_R^{\rm thin}$ & $39.67_{-0.72}^{+0.63}$ & $39.56_{-0.7}^{+0.66}$ & $39.27_{-0.62}^{+0.56}$ & $39.45_{-0.61}^{+0.67}$ & $39.23_{-0.6}^{+0.74}$ & $40.09_{-0.49}^{+0.59}$ & $31.2_{-0.14}^{+0.12}$ \\ \hline
$\sigma_z^{\rm thin}$ & $25.73_{-0.21}^{+0.21}$ & $25.72_{-0.25}^{+0.23}$ & $25.69_{-0.2}^{+0.22}$ & $25.67_{-0.23}^{+0.23}$ & $25.68_{-0.21}^{+0.25}$ & $25.77_{-0.22}^{+0.18}$ & $17.57_{-0.12}^{+0.13}$ \\ \hline
$\sigma_R^{\rm thick}$ & $42.43_{-1}^{+0.95}$ & $43.23_{-1.1}^{+0.96}$ & $42.98_{-0.73}^{+0.86}$ & $42.67_{-0.72}^{+0.96}$ & $43.51_{-0.82}^{+0.85}$ & $41.28_{-0.94}^{+0.71}$ & $48.51_{-0.6}^{+0.61}$ \\ \hline
$\sigma_z^{\rm thick}$ & $34.3_{-0.57}^{+0.51}$ & $34.48_{-0.53}^{+0.54}$ & $34.48_{-0.56}^{+0.58}$ & $34.66_{-0.55}^{+0.52}$ & $34.8_{-0.6}^{+0.55}$ & $33.8_{-0.55}^{+0.55}$ & $37.99_{-0.43}^{+0.41}$ \\ \hline
$\beta_R$ & $0.195_{-0.013}^{+0.011}$ & $0.192_{-0.013}^{+0.012}$ & $0.188_{-0.011}^{+0.01}$ & $0.192_{-0.012}^{+0.013}$ & $0.188_{-0.013}^{+0.013}$ & $0.2018_{-0.0093}^{+0.01}$ & \textcolor{magenta}{$0.01$} \\ \hline
$\beta_z$ & \textcolor{magenta}{$0.37$} & \textcolor{magenta}{$0.37$} & \textcolor{magenta}{$0.37$} & \textcolor{magenta}{$0.37$} & \textcolor{magenta}{$0.37$} & \textcolor{magenta}{$0.37$} & \textcolor{magenta}{$0.01$} \\ \hline
$1/R_{\sigma}^{\rm thin}$ & $0.073_{-0.003}^{+0.0037}$ & $0.0724_{-0.0031}^{+0.0031}$ & $0.0752_{-0.0034}^{+0.0034}$ & $0.0721_{-0.0026}^{+0.0028}$ & $0.0824_{-0.0031}^{+0.0038}$ & $0.0631_{-0.0027}^{+0.0029}$ & $0.0983_{-0.0024}^{+0.0034}$ \\ \hline
$1/R_{\sigma}^{\rm thick}$ & $0.1328_{-0.0051}^{+0.005}$ & $0.13_{-0.0046}^{+0.0056}$ & $0.1357_{-0.0045}^{+0.004}$ & $0.126_{-0.0048}^{+0.0035}$ & $0.1356_{-0.0044}^{+0.004}$ & $0.1319_{-0.0036}^{+0.005}$ & $0.1022_{-0.0034}^{+0.0034}$ \\ \hline
$\Theta_0$ & $231.9_{-1.5}^{+1.4}$ & $235.02_{-0.83}^{+0.86}$ & $223.3_{-1.4}^{+1.3}$ & $242.5_{-1.5}^{+1.6}$ & $249.8_{-1.5}^{+1.6}$ & $218.9_{-1.4}^{+1.5}$ & $237.3_{-1.6}^{+1.7}$ \\ \hline
$R_0$ & \textcolor{magenta}{$8$} & \textcolor{magenta}{$8$} & \textcolor{magenta}{$7.5$} & \textcolor{magenta}{$8.5$} & \textcolor{magenta}{$8$} & \textcolor{magenta}{$8$} & \textcolor{magenta}{$8$} \\ \hline
$\alpha_{z}$ & $0.0471_{-0.0019}^{+0.0016}$ & $0.0471_{-0.0019}^{+0.0019}$ & $0.0532_{-0.0017}^{+0.0017}$ & $0.0439_{-0.0017}^{+0.0016}$ & $0.0504_{-0.0018}^{+0.0018}$ & $0.0462_{-0.0018}^{+0.0016}$ & $0.0528_{-0.0019}^{+0.0019}$ \\ \hline
$\alpha_{R}$ & \textcolor{magenta}{$0$} & $0.67_{-0.26}^{+0.25}$ & \textcolor{magenta}{$0$} & \textcolor{magenta}{$0$} & \textcolor{magenta}{$0$} & \textcolor{magenta}{$0$} & \textcolor{magenta}{$0$} \\ \hline
\end{tabular}
\end{table*}
\subsection{Comparison with Bovy's kinematic model}
We carried out a more detailed analysis of the kinematic
model used by \citet{2012ApJ...759..131B}. We stress that there
are significant differences regarding both data and methodology between the
analysis done by us and by \citet{2012ApJ...759..131B}, and these should
be kept in mind when comparing the results.
Their sample is close to the plane $|b|<1.5^{\circ}$
and lies in the range $30^\circ<\ell<330^\circ$.
Being close to the plane, they cannot
measure vertical motion, but the advantage is they do not
have to worry about the dependence of asymmetric drift with vertical
height $z$.
The $\ell$ and $b$ range being different means that their data and ours
probe spatially different regions of the Milky Way.
If the disc is axisymmetric, we hope to get similar
answers, but not otherwise.

Their main analysis uses a single-population Gaussian model that does not
include an age-velocity dispersion relation, so we set $\beta_R \sim
\beta_{\phi} \sim \beta_{z} \sim 0$. They use a modified formula for the
asymmetric drift (Equation \ref{equ:stromberg_bovy}). In this formula we set
the parameter $k_{\rm ad}$ (in the notation of Bovy et al.\ $X$) to 0.85.
Our results are shown in column 1 of \tab{tbbovy}.  As mentioned earlier,
using RAVE data and a Gaussian model we obtain a negative value of
$R_\sigma^{\rm thin}=-34\kpc$ just as they do, in consequence of modelling a warm
population with a Gaussian model.  Our value of $\Theta_0$ is also in
agreement but our $\sigma_R$ is much larger than their value, $31.4\kms$.
Their sample could be dominated by cold stars on account of its proximity to the
plane.  They find $\sigma_{\phi}/\sigma_R=0.83$, which is higher by about 0.1
than our ratio for either RAVE or GCS using any type of model.

They also explored multiple populations with a prior on age given by an
exponentially declining star formation rate. However, they only quote
$\Theta_0,R_{0}$ and $\sigma_R$ for it.  For multiple populations, their prior on
age for the selected stars ignores the fact that scale height increases with
age. This will probably have little impact on $\Theta_0$, but their
$\sigma_R$ values cannot be compared with ours.  Also, they assume a priori
that $\beta_R=\beta_{\phi}=0.38$, but we have shown that $\sigma_R$ depends
upon the choice of $\beta_R$, and when we leave $\beta$ free, we obtain
values that differ from 0.38 (column 2 of \tab{tbbovy}).  If the thick disc
is included, the $\beta$ values are significantly reduced (column 3).  In
agreement with \citet{2012ApJ...759..131B}, we find that the value of
$\Theta_0$ is not affected much by the choice of age-velocity dispersion relation.
Including the thick disc leads to an increase in $\Theta_0$ by only 8 km
s$^{-1}$.  Interestingly, when $k_{\rm ad}$ is left free, we find the data favor
very high values (column 4).  This suggests that we are underestimating the
asymmetric drift, most probably due to our neglect of the vertical
dependence.

\subsection{Systematics} \label{sec:systematics}
Although we get quite precise values for most model parameters,
there are additional systematic uncertainties that we have
neglected. We performed some additional MCMC runs to investigate
these systematics. The results are summarized in \tab{shu_alpha}.
The first set of systematics is due to two parameters
that were kept fixed in our analysis, while the second set
is related to our choice of priors on the age and distance
distribution of stars.

The distance of the sun from the Galactic center $R_0$
and the radial gradient of circular speed $\alpha_R$ were
kept fixed at 8.0 kpc and zero for most of our analysis.
This is because these are strongly correlated with $\Theta_0$.
Using just the angular position and radial velocity of RAVE stars,
it is not possible to constrain them.
The effect of changing $R_0$ from 7.5 to 8.5 kpc can be seen
in column 3 and 4 of \tab{shu_alpha}, while the effect of changing
$\alpha_R$ from zero to $0.65 \kmskpc$ can be gauged by comparing
columns 1 and 2 in the same table. Using these tables, if needed
one can obtain values for any given $R_0$ and $\alpha_R$ by
linearly interpolating between the respective columns.
Increasing $\alpha_R$ increases $\Theta_0$, while the other parameters
are relatively unaffected.
Increasing $R_0$ increases $\alpha_z$ as well as $\Theta_0$.
Again, there is little change in other parameters.
The value of $\Omega_{\odot}$  was found to decrease from $30.8 \kmskpc$ at
$R_0=7.5$ kpc
to $29.4 \kmskpc$  at $R_0=8.5$ kpc. The above relationship tentatively
suggests that at $R_0 \sim 7.92$ one can match the
proper motion of Sgr A*.
We also checked the effect of setting $\alpha_z=0.0374$,
the value we expect from analytical models.
We found that this makes $\Theta_0 \sim 229.2 \kms$ and
$V_{\odot} \sim 8.0 $, which is not significantly far from
the value we get when  $\alpha_z$ is free.

We now discuss systematics related to our choice of priors.
Our main prior is that the age and distance distribution
of stars along a particular line of sight is
in accordance with the Besan\c{c}on model of the Galaxy.
Additionally, the distance distribution
for a given $I_{\rm DENIS}$ magnitude of a star depends upon the
isochrones that are used in the model.
As a crude way to gauge the sensitivity to our priors
in age, we
run a model with $\beta_z=\beta_R=0.01$ (column 7) which
makes the kinematics of the thin disc independent of age.
As expected, the thin and thick disc parameters change.
Other than this, $\alpha_z$ and $\Theta_0$ are found to
increase by 12\% and 2\% respectively.

Next, we test the effect of changing the distance prior.
This could be for example due to a systematic offset in
magnitudes predicted by the isochrones.
For this we alternately increase and
decrease our prior distance distribution by  multiplying the distances
by a factor of 1.1 and 0.9.
The values of $V_{\odot}$, $\Theta_0$ and $R_\sigma^{\rm thin}$ show
significant changes. It should be noted that this is only an
approximate way
to check the sensitivity of our results on the priors.
In reality, if magnitudes predicted by isochrones are systematically
wrong then the spatial density model that we use
will not match the number count of
stars obtained from photometric surveys. So, the mass density laws
of the model will have to be modified as well. The proper way to
do this is to do a dynamical modelling in which the
kinematics and the spatial distribution of stars
are fitted jointly to the observational data \citep[e.g.][]{2012MNRAS.426.1328B}.

The biggest source of systematic
uncertainty is related to the accuracy of the theoretical models
that we use. As discussed earlier in \sec{potmodel}, our
treatment of the vertical dependence of the
kinematics is not fully self consistent. In reality, for
a three dimensional system, the vertical and planar motions
are coupled to each other. To model such a system  properly
one needs a distribution
function that incorporates the third integral of motion.

Finally, our models will give rise to errors because they are kinematic
rather than dynamical models. Kinematic models offer greater freedom than
physics really allows. For example, the parameters $\sigma_R$ and
$\sigma_\phi$ of the Gaussian model are tightly coupled, as are $\beta_R$ and
$\beta_\phi$. The Shu model has fewer free parameters so is less open to this
criticism, but it fails to take into account the coupling between the
vertical profiles of $\sigma_R$ and the mean-streaming velocity
$\overline{v}_\phi$ \citep[e.g.][]{2012MNRAS.426.1324B}. It is not
unreasonable to hope that the values that emerge from the fits for
fundamentally superfluous parameters are similar to the values truly mandated
by physics, but noise in the data may confound this hope. Clearly, we should
proceed as quickly as possible to fitting RAVE with dynamical models like
those developed by \citet{2012MNRAS.426.1324B}.

\begin{table}
\caption{\label{tab:tbshufinal}
Shu model that best fits the RAVE data (same as column 6 \tab{tbshu}).
Quoted uncertainties  are purely
random and do not include systematics.}
\begin{tabular}{|l|l|} \hline
Model & RAVE SHU \\ \hline
$U_{\odot}$ & $10.96_{-0.13}^{+0.14} \kms $  \\ \hline
$V_{\odot}$ & $7.53_{-0.16}^{+0.16} \kms$  \\ \hline
$W_{\odot}$ & $7.539_{-0.09}^{+0.095} \kms$  \\ \hline
$\sigma_R^{\rm thin}$ & $39.67_{-0.72}^{+0.63} \kms$  \\ \hline
$\sigma_z^{\rm thin}$ & $25.73_{-0.21}^{+0.21} \kms$  \\ \hline
$\sigma_R^{\rm thick}$ & $42.43_{-1}^{+0.95} \kms$  \\ \hline
$\sigma_z^{\rm thick}$ & $34.3_{-0.57}^{+0.51} \kms$  \\ \hline
$\beta_R$ & $0.195_{-0.013}^{+0.011}$  \\ \hline
$\beta_z$ & \textcolor{magenta}{$0.37$}  \\ \hline
$1/R_{\sigma}^{\rm thin}$ & $0.073_{-0.003}^{+0.0037} {\rm
kpc}^{-1} $  \\ \hline
$1/R_{\sigma}^{\rm thick}$ & $0.1328_{-0.0051}^{+0.005} {\rm
kpc}^{-1}$  \\ \hline
$\Theta_0$ & $232.8_{-1.6}^{+1.7} \kms$   \\ \hline
$R_0$ & \textcolor{magenta}{$8$} kpc \\ \hline
$\alpha_{z}$ & $0.0471_{-0.0019}^{+0.0016}  $  \\ \hline
$\alpha_{R}$ & \textcolor{magenta}{$0.0 {\rm \ kpc}^{-1}$}   \\ \hline
\end{tabular}
\end{table}

\section{Summary and Conclusions}

In this paper, we have constrained the kinematic parameters of the Milky Way
disc using stars from the RAVE and the GCS surveys.  To constrain kinematic
parameters, we use analytic kinematic models based on the Gaussian and Shu
distribution functions.  We use these distribution functions, Padova stellar
tracks \citep{2008A&A...482..883M,1994AAS..106..275B} and the selection
functions of the surveys to predict the likelihood of each observed star.
For GCS data, which has full phase-space information for the stars, we
compute the likelihood in $({\bf x,v})$ phase space.  For RAVE data, we
choose to fit the likelihood in $(\ell,b,v_{\rm los})$ space to avoid use of
uncertain distances and proper motions.  We explored the full posterior
distribution of model parameters using the Markov Chain Monte Carlo
technique.
The parameters constrained include the Solar peculiar motion
($U_{\odot},V_{\odot},W_{\odot}$), the circular speed at the Sun $\Theta_0$,
a parameter $\alpha_z$ that controls the vertical gradient of
$R\partial\Phi/\partial R$, the age velocity
dispersion relations (via $\beta_{R,\phi,z}$, $\sigma_{R,\phi,z}^{\rm thin}$
and $\sigma_{R,\phi,z}^{\rm thick}$), and the scale lengths on which the
dispersions vary, $R_\sigma^{\rm thin}$ and $R_\sigma^{\rm thick}$.
Our results for both RAVE and GCS data are summarized in
Tables \ref{tab:tbgauss} and \ref{tab:tbshu}.
The final best fit model is given in \tab{tbshufinal}.

In our kinematic modelling the 
main assumption we make is that we assume a SFR, IMF and
density laws that describe the spatial distribution of stars in accordance
with the Besan\c{c}on model of \citet{2003A&A...409..523R}, but with slight
modifications as described in \citet{2011ApJ...730....3S}. 
So the kinematic results that we present are conditional upon the 
above assumption. 
Moreover, kinematic models offer greater
freedom than physics really allows.
So the accuracy of our kinematic results depends 
upon our ability to supply functional forms which are a good
approximation to the actual velocity distribution. 
To overcome these
concerns, one should fit both the kinematics
and the spatial distribution of stars together and they
should be dynamically linked via the potential in which the
stars move.

One could in principle constrain model parameters using the
two surveys, RAVE and GCS, simultaneously.
However, the two surveys probe different volumes, and it is
not clear that a single value of a given parameter, for example the solar
motion $U_\odot$, is appropriate for both volumes: the immediate vicinity of
the Sun may be moving with respect to the wider disc, for example. If such
systematic differences exist, the simple models we are fitting cannot provide
an adequate account of the entire body of data, and parameter values obtained
from a joint fit will be of doubtful physical significance.  Hence, in this
paper, we first analyzed the surveys separately and tried to understand the
systematics.  Then, having understood the extent to which each survey
constrained each parameter, we fixed values of some parameters from the
results of one survey while analysing the other.  We do this only for those
parameters which we believe should take the same values for both surveys.

The Gaussian model proves to be unsuitable for estimating disc parameters
such as $R_\sigma^{\rm thin}$ and $V_{\odot}$ because the fits prove to be
strongly degenerate.  The Gaussian model gives different values of
$R_\sigma^{\rm thin}$ for RAVE and GCS.  For RAVE it predicts negative
values, implying that $\sigma_R$ increases outwards.  This result is
inconsistent with the disc's scale height and value of $\sigma_z/\sigma_R$
being constant.  Negative values of $R_\sigma^{\rm thin}$ also disagree with
the findings of \citet{1989AJ.....97..139L}.  The Shu model has three fewer
parameters than the Gaussian model and this helps it to break the degeneracy
between $R_\sigma^{\rm thin}$ and $V_{\odot}$. It gives positive and consistent
values for $R_\sigma^{\rm thin}$ for both RAVE and GCS.  The Shu model also fits
the RAVE data better than the Gaussian model, especially with regard to
stars' values of $v_\phi$.

The RAVE data allow us to constrain the Solar peculiar motion and the local
circular speed quite precisely.  Our $U_{\odot}$ and $W_{\odot}$ are in good
agreement with the results of \citet{2010MNRAS.403.1829S}, but our
$V_{\odot}$ is lower by 5 km s$^{-1}$.  The RAVE $U_{\odot}$ and $W_{\odot}$
are within 2-$\sigma$ range of GCS values, but $V_{\odot}$ is lower by $2
\kms$.  Using $R_0=8.0$ kpc and assuming $\partial v_{\rm c}/\partial R=0$
we get $\Theta_0 \sim 232$ km s$^{-1}$.  Combining the estimates of $\Theta_0$
and $V_{\odot}$, we find the Solar angular velocity with respect to the
Galactic Center to be in good agreement with the measured proper motion of
Sgr A*. We find that if the fall of mean azimuthal velocity
with height $z$ above the mid plane is neglected then this
leads to an underestimation of $\Theta_0$.

Although our random uncertainty regarding most parameters is quite small,
due to large number of stars in the RAVE survey, significant
sources of systematic uncertainty remain, especially
regarding $\Theta_0$ and $V_{\odot}$.
Our treatment of the vertical dependence of
the kinematics is not fully self consistent. This needs to
be investigated with models that can handle the third
integral of motion, e.g., models based on action
integrals.
Also we need to explore dynamical models that are self-consistent
rather than pure kinematic models as studied here.
The values of
$\Theta_0$ and $V_{\odot}$ are also sensitive
to the priors on age and distance distribution of stars. So systematic errors
of the order of the uncertainty in the priors are
also expected.

When using the Shu model, all parameters except $V_{\odot}$ and
thick-disc parameters, show similar values for RAVE and GCS.  Since there are
very few thick disc stars in GCS, we deem the RAVE thick disc parameters to
be more reliable.  Also, the uncertainty on $R_\sigma^{\rm thin}$ and
$R_\sigma^{\rm
thick}$ is substantially less for RAVE than for GCS.  The only parameter that
is constrained better by GCS than RAVE is $\beta_z$ and this is partly due to
the fact that we only use radial velocities in RAVE.  In an attempt to build
a concordance model, and to enable better comparison between the two data
sets, we fix $\beta_z$ in RAVE to GCS values and then fix $R_\sigma^{\rm thin}$ and
$R_\sigma^{\rm thick}$ in GCS to RAVE values.  Doing so we find that RAVE results
are within 3$\sigma$ of GCS results.  The most significant difference between
the two is the value of $V_{\odot}$, which is lower for RAVE by about $2\kms$.  The presence of prominent kinematic substructures in GCS could be
responsible for this discrepancy. However, inaccuracy in
our vertical treatment of kinematics could also be responsible.

We find that the age-velocity dispersion relations in general satisfy
$\beta_R < \beta_{\phi}<\beta_z$, with $\beta_{\phi}$ is closer to $\beta_R$
than $\beta_z$, contrary to the finding of \citet{2009MNRAS.397.1286A}. This
result is consistent with the physical principle that peculiar motions in the
radial and azimuthal directions are strongly coupled by epicyclic dynamics,
and large decoupled from vertical motions.  The fitted $\beta$ values depend
on whether the thick disc is added separately or is left to be represented by
the old tail of the thin disc, and they are naturally higher when it is not
added separately.  The axial ratio $\sigma_z/\sigma_R$ of the thin disc
velocity ellipsoid for the 10 Gyr population is consistent with those
predicted by \citet{2008ASPC..396..241S} for cloud scattering.  Our values of
$\beta_R$ and $\beta_z$ agree well with age-velocity profiles measured by
\citet{2013A&A...558A...9M} for ages $\la7\Gyr$ in simulations of disc
galaxies.  At ages larger than $7\Gyr$, a model that consists of power-law
growth in the thin disc combined with a distinct thick-disc population, is
too crude to represent the simulations adequately. In future it may be
appropriate to use more elaborate models inspired by simulations.

In the Shu model, the thick disc velocity dispersions for $R_g=R_0$ are very
similar to those of the old thin disc.  However, the radial scale length of
the thick-disc velocity dispersions, $R_\sigma^{\rm thick}$, proved to be
much smaller than that of the thin disc.  \citet{2012ApJ...753..148B}
suggested a decrease of radial density scale length with age.  In this
regard, the role of our adopted priors on age and distance distribution of
stars needs to be investigated further.

Given the essential role that age plays in disc dynamics, it is unfortunate
that the ages of stars are so hard to measure. Fortunately, big advances in
this area are expected soon.  Stellar astroseismology with missions like
CoRoT and KEPLER makes it possible to measure ages more accurately than
before \citep{2010ApJ...713L.169C,2011Sci...332..213C,2008A&A...488..705A},
and Gaia will dramatically improve age estimates by geometrically determining
distances to large numbers of stars. Meanwhile, chemical abundances,
especially of the alpha elements, provides a fair proxy for age at a given
metallicity.  Hence, studying the relationship of kinematic properties with
abundance will be crucial.  \citet{2012ApJ...753..148B} argued that each
mono-abundance population has a distinct spatial distribution, and we expect
cohorts of coeval stars to have spatial distributions that are characteristic
of their ages.


\section*{Acknowledgments}
SS is funded through ARC DP grant 120104562
(PI Bland-Hawthorn) which supports the HERMES project.
JBH is funded through a Federation Fellowship from the Australian
Research Council (ARC).

Funding for RAVE has been provided by: the Australian
Astronomical Observatory; the Leibniz-Institut fuer Astrophysik
Potsdam (AIP); the Australian National University;
the Australian Research Council; the French National Research
Agency; the German Research Foundation (SPP 1177
and SFB 881); the European Research Council (ERC-StG
240271 Galactica); the Istituto Nazionale di Astrofisica at
Padova; The Johns Hopkins University; the National Science
Foundation of the USA (AST-0908326); the W. M.
Keck foundation; the Macquarie University; the Netherlands
Research School for Astronomy; the Natural Sciences
and Engineering Research Council of Canada; the Slovenian
Research Agency; the Swiss National Science Foundation;
the Science \& Technology Facilities Council of the UK;
Opticon; Strasbourg Observatory; and the Universities of
Groningen, Heidelberg and Sydney. The RAVE web site is
at http://www.rave-survey.org.

\end{document}